\documentclass[fleqn,usenatbib]{mnras}

\usepackage{newtxtext,newtxmath}
\usepackage{placeins}
\usepackage[T1]{fontenc}

\DeclareRobustCommand{\VAN}[3]{#2}
\let\VANthebibliography\thebibliography
\def\thebibliography{\DeclareRobustCommand{\VAN}[3]{##3}\VANthebibliography}

\usepackage{footnote}
\usepackage{graphicx}	
\usepackage{amsmath}	

\title[Low Luminosity Supernovae IIP]{Low luminosity Type II supernovae – IV. SN 2020cxd and SN 2021aai, at the edges of the sub--luminous supernovae class}

\author[G. Valerin et al.]{
G. Valerin,$^{1,2}$\thanks{E--mail: giorgio.valerin@studenti.unipd.it}
M. L. Pumo$^{1,3,4}$,
A. Pastorello$^{1}$,
A. Reguitti$^{5,6,1}$, 
N. Elias--Rosa$^{1,7}$, \newauthor
C. P. G\'{u}tierrez$^{8,9}$, 
E. Kankare$^{9}$, 
M. Fraser$^{10}$,
P. A. Mazzali$^{11,12}$,
D. A. Howell$^{13,14}$, \newauthor
R. Kotak$^{9}$, 
L. Galbany$^{7,15}$, 
S. C. Williams$^{8,9}$, 
Y.--Z. Cai$^{16}$,
I. Salmaso$^{1,2}$, \newauthor
V. Pinter$^{17,18}$,
T. E. M\"{u}ller-Bravo$^{7}$, 
J. Burke$^{13,14}$, 
E. Padilla Gonzalez$^{13,14}$, \newauthor
D. Hiramatsu$^{19}$, 
C. McCully$^{13}$, 
M. Newsome$^{13,14}$, 
C. Pellegrino$^{13,14}$
\\
\\
$^{1}$INAF -- Osservatorio Astronomico di Padova, Vicolo dell’Osservatorio 5, I-35122 Padova, Italy\\
$^{2}$Dipartimento di Fisica e Astronomia ``G. Galilei'', Universit\`{a} degli studi di Padova Vicolo dell’Osservatorio 3, I-35122 Padova, Italy \\
$^{3}$Dipartimento di Fisica e Astronomia ``E. Majorana'', Universit\`{a} degli studi di Catania, Via S. Sofia 64, I-95123 Catania, Italy \\
$^{4}$Laboratori Nazionali del Sud – INFN, Via S. Sofia 64, I-95123 Catania, Italy \\
$^{5}$Departamento de Ciencias F\'{i}sicas – Universidad Andres Bello, Avda. Rep\'{u}blica 252, 8320000, Santiago, Chile\\
$^{6}$Millennium Institute of Astrophysics, Nuncio Monsenor S\'{o}tero Sanz 100, Providencia, 8320000, Santiago, Chile\\
$^{7}$Institute of Space Sciences (ICE, CSIC), Campus UAB, Carrer de Can Magrans s/n, E-08193 Barcelona, Spain \\
$^{8}$Finnish Centre for Astronomy with ESO (FINCA), FI-20014 University of Turku, Finland\\
$^{9}$Tuorla Observatory, Department of Physics and Astronomy, University of Turku, 20014, Turku, Finland\\
$^{10}$School of Physics, University College Dublin, Belfield, Dublin 4, Ireland\\
$^{11}$Astrophysics Research Institute, Liverpool John Moores University, IC2, Liverpool Science Park, 146 Brownlow Hill, Liverpool L3 5RF, UK\\
$^{12}$Max-Planck-Institut f\"{u}r Astrophysik, Karl-Schwarzschild Str. 1, D-85748 Garching, Germany \\
$^{13}$Las Cumbres Observatory, 6740 Cortona Dr. Suite 102, Goleta, CA, 93117, USA\\
$^{14}$Department of Physics, University of California, Santa Barbara, Santa Barbara, CA, 93106, USA\\
$^{15}$Institut d’Estudis Espacials de Catalunya (IEEC), E-08034 
Barcelona, Spain. \\
$^{16}$Physics Department and Tsinghua Centre for Astrophysics (THCA), Tsinghua University, Beijing 100084, China \\
$^{17}$Isaac Newton Group (ING), Apt. de correos 321, E-38700, Santa Cruz de La Palma, Canary Islands, Spain \\
$^{18}$University of Craiova, Str. A. I. Cuza nr. 13, Craiova, 200585, Romania \\
$^{19}$ Center for Astrophysics, Harvard \& Smithsonian, Cambridge, Massachusetts, MA 02138, US
}

\date{Accepted XXX. Received YYY; in original form ZZZ}

\pubyear{2022}

\begin{document}
\label{firstpage}
\pagerange{\pageref{firstpage}--\pageref{lastpage}}
\maketitle

\begin{abstract}
Photometric and spectroscopic data for two Low Luminosity Type IIP Supernovae (LL SNe IIP) are presented.
SN 2020cxd reaches a peak absolute magnitude $M_{r}$ = --13.90 $\pm$ 0.05 mag two days after explosion, subsequently settling on a plateau for $\sim$120 days. Through the luminosity of the late light curve tail, we infer a synthesized \textsuperscript{56}Ni mass of (1.8$\pm$0.5) $\times$ 10$^{-3}$ M$_{\odot}$.
During the early evolutionary phases, optical spectra show a blue continuum ($T$ $>$ 8000 K) with broad Balmer lines displaying a P Cygni profile, while at later phases Ca II, Fe II, Sc II and Ba II lines dominate the spectra.
Hydrodynamical modelling of the observables yields $R$ $\simeq$ 575 $R_{\odot}$ for the progenitor star, with $M_{ej}$ = 7.5 M$_{\odot}$ and $E$ $\simeq$ 0.097 foe emitted during the explosion. This low--energy event originating from a low--mass progenitor star is compatible with both the explosion of a red supergiant (RSG) star and with an Electron Capture Supernova arising from a super asymptotic giant branch star.
SN 2021aai reaches a maximum luminosity of $M_{r}$ = --16.4 mag (correcting for $A_{V}$=1.9 mag), and displays a remarkably long plateau ($\sim$140 days).
The estimated \textsuperscript{56}Ni mass is (1.4$\pm$0.5) $\times$ 10$^{-2}$ M$_{\odot}$.
The expansion velocities are compatible with those of other LL SNe IIP (few 10$^{3}$ km s$^{-1}$).
The physical parameters obtained through hydrodynamical modelling are $R$ $\simeq$ 575 R$_{\odot}$, $M_{ej}$ = 15.5 M$_{\odot}$ and $E$ = 0.4 foe. SN 2021aai is therefore interpreted as the explosion of a RSG, with properties that bridge the class of LL SNe IIP with standard SN IIP events.
\end{abstract}

\begin{keywords}
supernovae: general -- supernovae: individual: SN 2020cxd, SN 2021aai
\end{keywords}



\section{Introduction}

During the last two decades, the transient Universe has been inspected in an unprecedented fashion thanks to new instruments and dedicated surveys: therefore, the discovery of new classes of transients did not come as a surprise.
In particular, the so called "luminosity gap" \citep[][]{Kasliwal2012} separating Classical Novae ($M_{V}$ $\sim$ --10 mag) and standard type II Supernovae (SNe; $M_{V}$ $\sim$ --15 mag) has been populated with several peculiar phenomena.
Among the “gap transients” (see e.g. \citealt{PastoMorgan2019}) can be identified stellar mergers (Luminous Red Novae), stellar eruptions (Luminous Blue Variables) and even authentic, though weak, Core--Collapse SNe. In particular, faint SNe explosions are expected to be produced when the sub--energetic explosion of a very massive stars leads the inner stellar mantle to fall back onto the core \citep{PumoModel2017}. These SNe are characterized by the ejection of tiny \textsuperscript{56}Ni amounts (e.g. \citealt{Moriya2010}). The collapse of an O--Ne--Mg core of a moderate--mass super--AGB star is also expected to produce faint transients known as electron--capture SNe (ECSNe) (e.g. \citealt{NomotoECSN,RitossaECSN,Kitaura2006,PoelrandsSingleECSN}), although there is no consensus yet on whether we already witnessed such an explosion. Given their faintness and low synthesised  \textsuperscript{56}Ni mass, the so--called Intermediate--Luminosity Red Transients (ILRTs; \citealt{Botticella2008S,MaxILRT,Yongzhi2021}) are considered to be among the most promising candidates. The electron--capture SN scenario, however, can potentially produce transients with different observable properties. The peculiar type II SN 2018zd \citep{Hiramatsu2018zd} has shown a remarkable compatibility with the ECSN scenario, although no consensus has been reached yet on the nature of this object \citep{Zhang2018zd,Callis2018zd}.

Together with this array of unusual and little studied transients, there is a group of Low Luminosity SNe type IIP (LL SNe IIP) lying towards the upper end of the "luminosity gap". 
The first identified object of this class was SN 1997D \citep{Turatto1998,Benetti1997D}, which was reported as one of the faintest SN observed to that date, peaking at M$_{B}$=--14.65 mag. The late time decline was also unusually faint, compatible with the ejection of just 2$\times$10$^{-3}$ M$_{\odot}$ of $\textsuperscript{56}$Ni, one order of magnitude lower than the typical value for standard SNe IIP (a few 10$^{-2}$ M$_{\odot}$, \citealt{AndersonAverageIIP}). The first scenario proposed to explain this event envisioned a massive progenitor (25--40 M$_{\odot}$), and the fallback on the black hole formed during the collapse would account for the low amount of energy emitted \citep{ZampieriBH1998,ZampPrel2003}. 
Important steps towards understanding the nature of LL SNe IIP progenitors were taken thanks to observational studies on samples of standard type IIP SNe \citep{Smartt2009,Smartt2015}, which determined that the progenitor stars of SNe IIP were Red Super Giants (RSGs) with low Zero Age Main Sequence (ZAMS) masses between 8 and 18 M$_{\odot}$. These findings disfavoured the scenario of the massive progenitor for LL SNe IIP \citep{Eldridge2005cs,Fraser2009md,CrockettProgenitor}. 
This study was based on the direct detections of the progenitor star in archival images before the SN explosion, and subsequent matching with theoretical evolutionary tracks.
A different approach to determine the progenitor mass consists in computing hydrodynamical models to describe observed lights curves and expansion velocities (e.g.  \citealt{Bersten11,ChugaiandPasto2007,UtrobinAndChugai2008,PumoEZampieriModel,Lisakov18,Martinezb,Kozyreva2021}). The mass estimates obtained with this method are sometimes higher (14--18 M$_{\odot}$) than the ones obtained through direct progenitor detection, possibly due to an overestimate of the ejected mass due to spherical symmetry approximation \citep{ChugaiOverestimate2009}.
There has been also a third approach \citep{FranssonOI,Maguire2010,Jerk2012,Jerk2014,Jerk2018, Lisakov17, Lisakov18, Dessart2021}: the nebular [OI] doublet $\lambda \lambda$ 6300,6364 observed in the late--time spectra is used as a tracer of the core mass of the progenitor star and hence of its ZAMS mass. 

The method described above was developed to study standard SNe IIP, but it was also applied to LL SNe IIP, when possible: spectral modelling results are so far consistent with the lack of massive progenitors ($\sim$ 20 M$_{\odot}$) for LL SNe IIP \citep{ClaudiaGutierrez2020synthesis}.
Studies on the photometric and spectroscopic evolution of larger samples (up to 15 objects) of LL SNe IIP \citep{PastoLLSNe2004,SpiroPasto2014} found that these transients share strikingly similar features. The light curves of
LL SNe IIP are characterized by a quick rise to maximum (few days), followed by a plateau lasting $\sim$ 100 days, before finally settling on a late time linear decay compatible with the ejection of a small amount of \textsuperscript{56}Ni ($<$10$^{-2}$ M$_{\odot}$). The temperature evolution is quite homogeneous among the various objects observed, with a rapid cooling at early phases leading to a temperature of 10$^{4}$ K at 10 days, and a slower subsequent decline (6000--8000 K at 30 days). The expansion velocities inferred from the spectral lines also show a fast decrease from some 10$^{3}$ km s$^{-1}$ in the first week after explosion to $\sim$ 2000 km s$^{-1}$ one month after.
These findings are consistent with those inferred for standard SNe IIP: transients with dimmer plateaus show lower expansion velocities and eject less \textsuperscript{56}Ni \citep{Hamuy2003,Gutierrez2017b}. \citet{PastoLLSNe2004} and \citet{SpiroPasto2014} proposed that LL SNe IIP are the least energetic end of the continuous distribution of SNe IIP in the parameter space (progenitor mass, plateau luminosity, \textsuperscript{56}Ni synthesised, expansion velocities). This proposition is corroborated by the presence of "transitional" objects, showing intermediate properties between LL SNe IIP and standard SNe IIP, like SN 2009N \citep{2009N} and SN 2008in \citep{SN2008in}. Furthermore, \cite{PumoModel2017} show that the parameter E / M$_{ej}$ ``guide'' the distribution of the SNe IIP class in the parameters space, where LL SNe IIP form the underluminous tail.

In the context of LL SNe IIP, we present photometric and spectroscopic data that we collected for two objects belonging to this class: SN 2020cxd\footnote{SN 2020cxd has been the studied by \cite{Sheng2020cxd}. Here we provide additional photometric and spectroscopic coverage of this target. Just before our submission, \cite{Kozyreva2022_cxd} presented an additional paper on the modelling of 2020cxd.}, one of the faintest LL SNe IIP observed to date, and SN 2021aai, which belongs to the brighter end of the class. In Sect. \ref{DataReduction} we discuss the methodology used to obtain and reduce the data, while in Sect. \ref{DiscoveryPhotometry} the photometric data are presented. In Sect. \ref{Spectroscopic} we analyse the spectra and in Sect. \ref{BlackBodyF} we discuss the physical parameters obtained through blackbody fits. In Sect. \ref{NiEst} we estimate the \textsuperscript{56}Ni ejected mass during the explosion and compare the results with similar objects. In Sect. \ref{Hydro} we perform hydrodynamic modeling on our targets in order to infer information on their progenitor stars. Finally, in Sect. \ref{Concl} we summarise the results obtained.

\begin{figure*}
\centering
\begin{minipage}[b]{.45\textwidth}
\includegraphics[width=1\columnwidth]{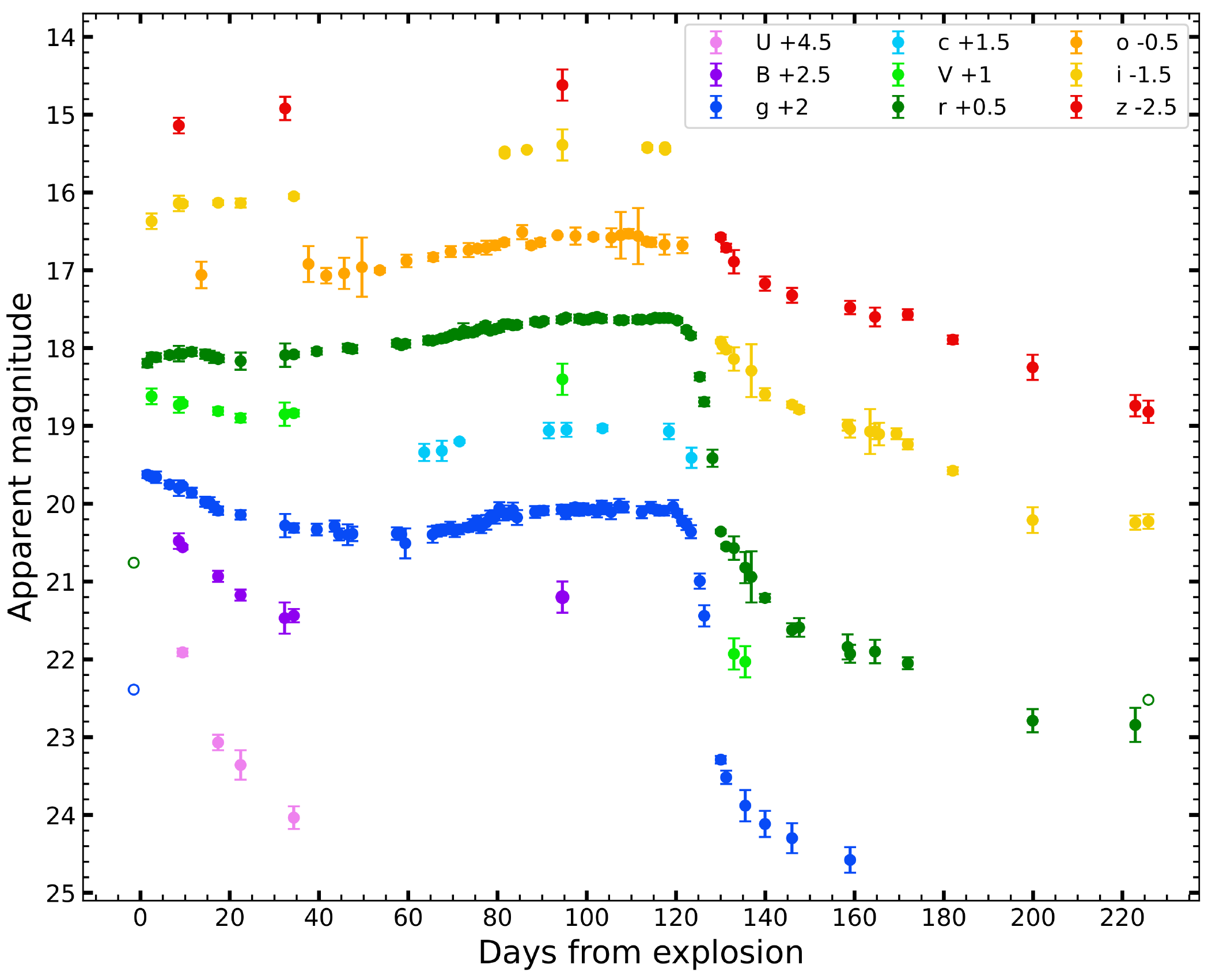}
\caption{Optical light--curves of SN 2020cxd. Empty circles represent upper magnitude limits.  }
\label{SN2020cxd_Phot}
\end{minipage}\qquad
\begin{minipage}[b]{.45\textwidth}
\includegraphics[width=1\columnwidth]{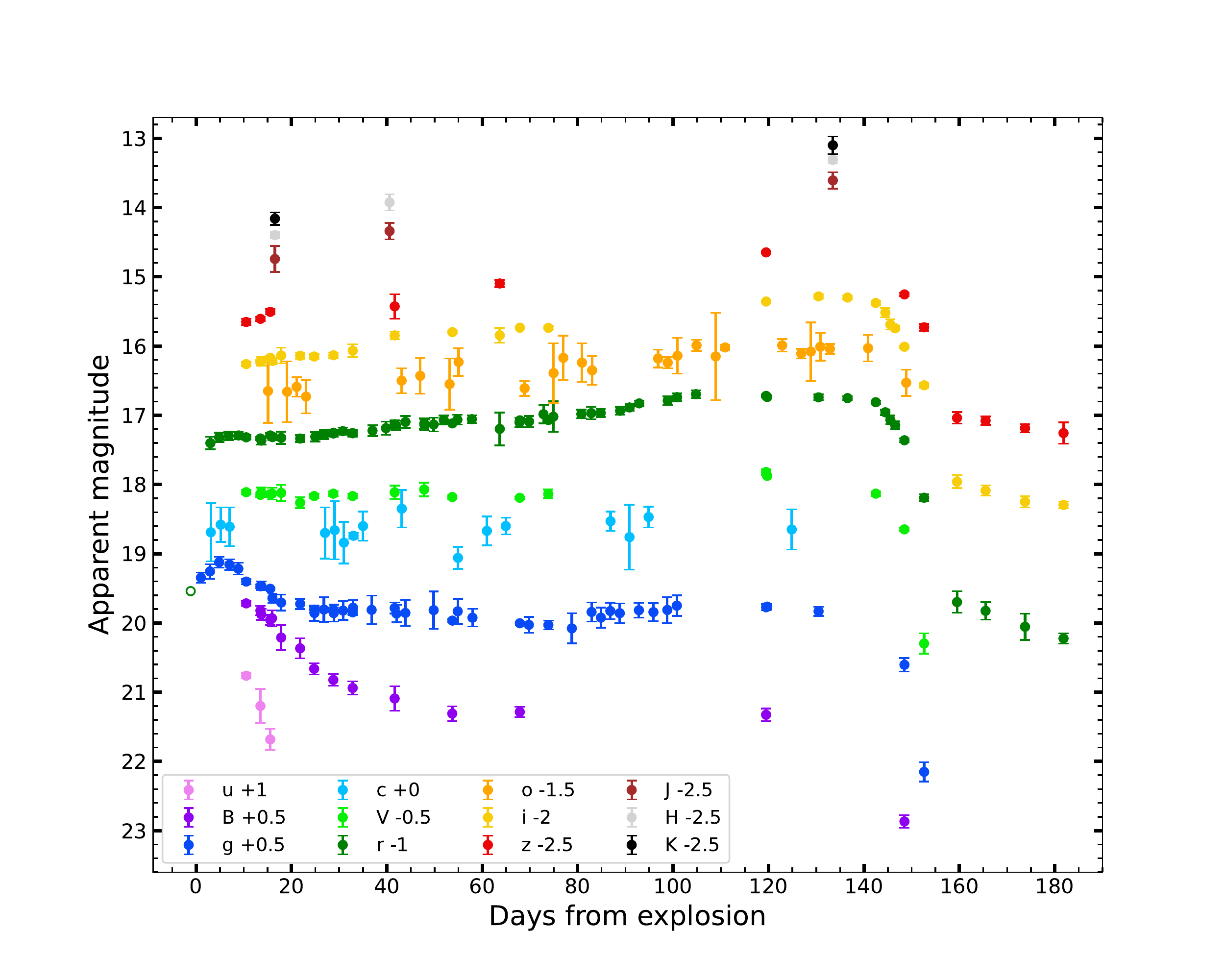}
\caption{Optical and NIR light curves of SN 2021aai. Empty circles represent upper magnitude limits.}
\label{SN2021aai_Phot}
\end{minipage}
\end{figure*}

\section{Data reduction} \label{DataReduction}
The objects in this paper were followed with several instruments at different facilities, whose details are reported in Table \ref{Instrum}. In particular, the majority of the private data we present in this work was collected with the Nordic Optical Telescope (NOT) within the NOT Unbiased Transient Survey 2 (NUTS2) collaboration \citep{Holmbo2019ATEL}, with the Liverpool telescope and within the Global Supernova Project \citep{AndyLCO}.
Image reduction was performed through standard \texttt{IRAF} tasks \citep{IRAF_Toby}, removing the overscan and then correcting for bias and flat field. When multiple exposures were taken on the same night, we combined them to improve the signal to noise (S/N) ratio. To measure the observed magnitudes of our targets, we used a dedicated, \texttt{python} based pipeline called \texttt{SNOoPy}\footnote{A detailed package description can be found at http://sngroup.oapd.inaf.it/snoopy.html.} \citep{snoopy}. \texttt{SNOoPy} consists in a series of Python scripts calling \texttt{IRAF} standard tasks like \texttt{DAOPHOT} through \texttt{PYRAF}, and it was designed for Point Spread Function (PSF) fitting of multi--wavelength data acquired from different instruments and telescopes. The PSF model was built from the profiles of isolated, unsaturated stars in the field. The instrumental magnitude of the transient was then retrieved by fitting this PSF model and accounting for the background contribution around the target position through a low--order polynomial fit. The error on this procedure was obtained through artificial stars created close to the target, with magnitudes and profiles coincident with those inferred for the object. The dispersion of the artificial stars instrumental magnitudes was combined in quadrature with the PSF fitting error given by \texttt{DAOPHOT} to obtain the total error associated with that measure. Zero Point (ZP) and Colour Term (CT) corrections were computed for each instrument by observing standard fields: SDSS \citep{SDSS} was used as reference for Sloan filters, the \cite{Landolt} catalogue was used for Johnson filters and the 2MASS \citep{2MASS} catalogue was used for Near Infrared (NIR) filters. It is worth noticing that in the NIR we assumed negligible CT, so we only computed the ZP correction. In order to account for non--photometric nights, we selected a series of stars in the field of each transient: measuring the average magnitude variation of the reference stars with respect to the catalogued magnitudes, we computed the ZP correction for each night in each filter. Applying ZP and CT corrections to the instrumental magnitudes of our targets, we obtained the apparent magnitudes reported in this paper. We adopted the AB magnitudes system for $ugriz$ bands and Vega magnitudes for $BVJHK$ bands.
For the "Asteroid Terrestrial--impact Last Alert System" (ATLAS) data \citep{TonryATLAS}, we combined the flux values obtained through forced photometry reported in their archive\footnote{https://fallingstar--data.com/forcedphot/}, and converted the result into magnitudes as prescribed in the ATLAS webpage. The photometric measurements we obtained are reported in Appendix \ref{data_tables} (full tables are available in the online supplementary material).

The original spectra presented in this work (see Table \ref{LogSpectra}) were reduced through standard \texttt{IRAF} routines contained in the package \texttt{CTIOSLIT}.
All spectra were corrected for bias and flat-field before extracting the 1-D spectrum. Sky lines and cosmic rays were removed, wavelength and flux calibrations were applied using arc lamps and spectrophotometric standard stars. Finally, spectra were corrected for telluric lines, they were flux calibrated an additional time on the broad--band photometric data obtained at the same phase, and they were corrected for redshift and reddening (discussed in Sect. \ref{DiscoveryPhotometry}).
In particular, spectra taken with  the NOT were reduced through the \texttt{ALFOSCGUI}\footnote{More details at https://sngroup.oapd.inaf.it/foscgui.html} pipeline \citep{snoopy}, specifically designed to reduce spectra within the NUTS2 collaboration.   
The spectra presented in this article will be available on the WISeREP repository \citep{WiseRep}.




\section{Discovery and photometric evolution} \label{DiscoveryPhotometry}
\subsection{SN 2020cxd photometric properties}
SN 2020cxd is a LL SN IIP discovered on 2020 Feb 19 \citep{2020cxd_Discovery} at the coordinates RA = $17^{h} 26^{m} 29^{s}.26$ Dec = +71\textdegree 05' 38''.58 in the spiral galaxy NGC 6395, classified as Scd \citep{DeVaucouleurs} and with a redshift z = 0.003883 $\pm$ 0.000002 \citep{Redshift2020cxd}. As noticed by \cite{Sheng2020cxd}, the distance measurements for the host galaxy vary between 19 and 23 Mpc, depending on the methodology used. In this paper we adopt a distance modulus of $\mu$ = 31.60 $\pm$ 0.20 mag (or 20.9 $\pm$ 1.4 Mpc), obtained by averaging the six different estimates obtained through the Tully--Fisher method and reported on the NED database \citep{cxdDist1,Distance2021aai,cxdDist3,cxdDist4,cxdDist5,cxdDist6}. We assumed a cosmology where H$_{0}$ = 73 km s$^{-1}$ Mpc$^{-1}$, $\Omega_{\Lambda}$ = 0.73 and $\Omega_{M}$ = 0.27 \citep{SpergelCosmology}, which will be used throughout this work. 
The Galactic absorption in the direction of NGC 6395 is $A_{V}$ = 0.11 $\pm$ 0.03 mag, from \cite{SchlaflyNEDReddening2011}, under the assumption that R$_{V}$=3.1 (\citealt{Cardelli3.1}; which will be used throughout this work). Early spectra do not show evidence of the interstellar Na I D absorption doublet at the host galaxy redshift, allowing us to estimate as negligible the absorption along the line of sight (see Sect. \ref{Spectroscopic}).


In Figure \ref{SN2020cxd_Phot}, we report the multi--wavelength photometry of SN 2020cxd collected up to 230 days after explosion. The early rise in luminosity was not observed, since the object was first detected when it was already on the plateau. However, thanks to a deep upper limit ($r$ $>$ 20.3 mag) obtained just three days before the discovery by the Zwicky Transient Facility (ZTF; \citealt{ZTF}), it is possible to constrain the explosion epoch with small uncertainty to MJD = 58897.0 $\pm$ 1.5. Even on the plateau, the brightness was not strictly constant: at first there was a decline, with the transient dimming from $M_{r}$ = --14.13 mag at 10 days to $M_{r}$ = --14.00 mag at 22 days (typical photometric error of 0.04 mag). This luminosity decrease was more marked in the blue bands. This behaviour is clear in the $g$ band, where the absolute magnitude declined from $M_{g}$ = --13.97 mag to $M_{g}$ = --13.20 mag in the first 60 days. Thereafter, the brightness consistently increased to $M_{g}$ = --13.58 mag and $M_{r}$ = --14.48 mag before finally fading from the plateau at $\sim$120 days. \cite{ClaudiaGutierrez2020synthesis} attributed the different behaviour of the $g$ band compared to the $r$ band to the shift of the Spectral Energy Distribution (SED) peak from the ultraviolet (UV) to the optical domain.
The drop from the plateau was very sharp, with the object fading by 2.9 mag in the $r$ band and 3.2 mag in the $g$ band in just 10 days. Finally, the luminosity evolution settled on a linear decline powered by the \textsuperscript{56}Ni synthesised during the explosion. More details in Sect. \ref{Ni56}.





\begin{figure}
\includegraphics[width=1\columnwidth]{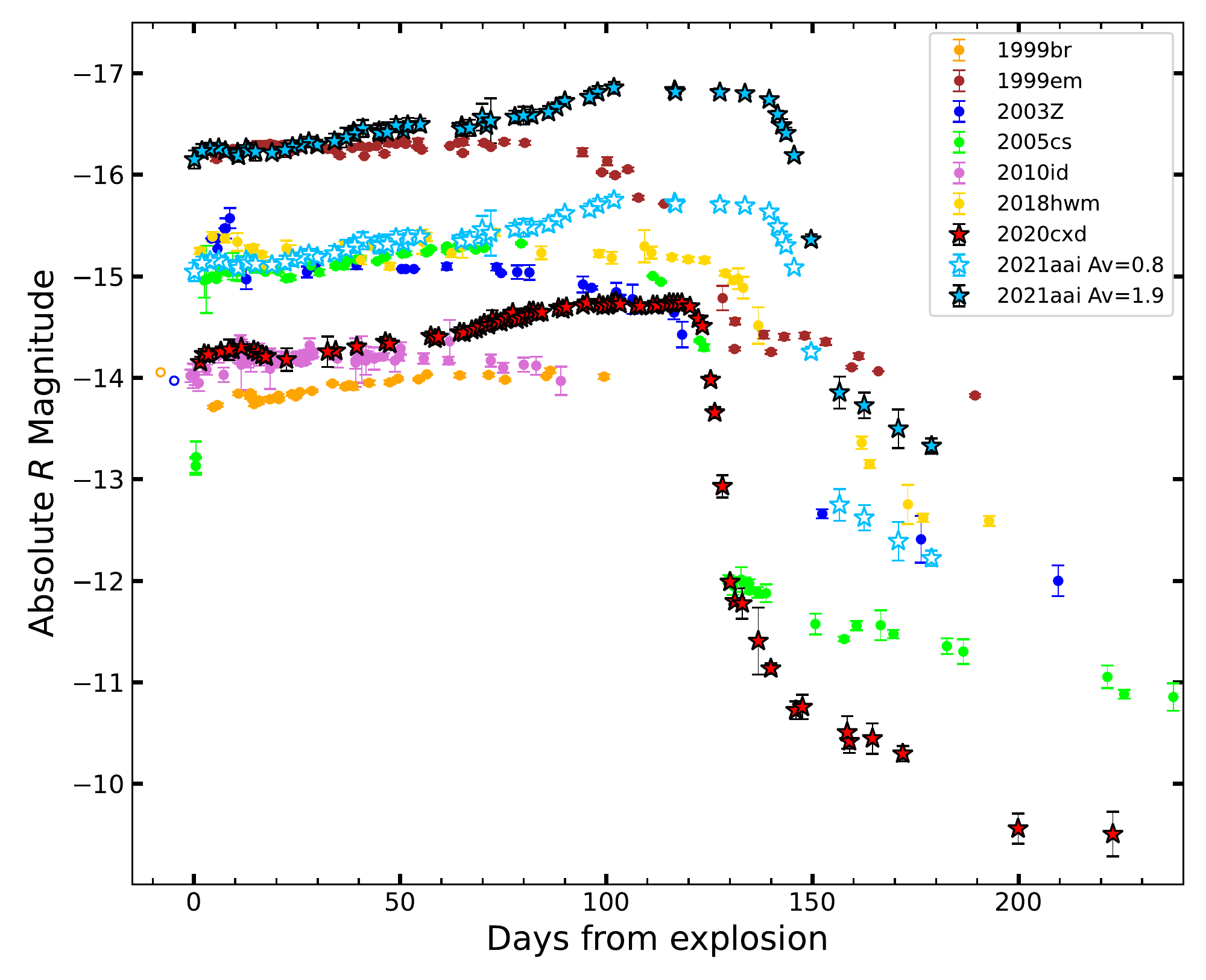}
\caption{Comparison of the $R$ band evolution for a sample of SNe IIP, spanning from some of the faintest objects observed, like SN 1999br, up to events like SN 1999em, which are close to the standard SNe IIP. }
\label{Comparison}
\end{figure}

\subsection{SN 2021aai photometric properties} \label{r-R}
SN 2021aai was discovered at the coordinates RA = $07^{h} 14^{m} 26^{s}.86$ Dec = +84\textdegree 22' 51''.46 on 2021 Jan 12 \citep{Discovery2021aai} in NGC2268, an SAB(r)bc \citep{DeVaucouleurs} at a redshift of z = 0.007428 $\pm$ 0.000007 \citep{Redshift2020cxd}. We adopt a distance modulus of $\mu$ = 32.47 $\pm$ 0.20 mag (31.2 $\pm$ 1.7 Mpc) obtained through one of the most recent Tully--Fisher estimates \citep{Distance2021aai}. According to \cite{SchlaflyNEDReddening2011}, the reddening internal to the Milky Way along the line of sight towards NGC2268 is $A_{V}$ = 0.170 $\pm$ 0.003 mag. Unlike in SN 2020cxd, the Na I D absorption doublet was detected in the first two spectra obtained (see Sect. \ref{Spectroscopic}), with an Equivalent Width (EW) of 1.6 \AA. Some relationships between reddening and Na I D EW typically saturate with such high values of EW \citep{Poznanski2012Na}, so we estimate a lower limit to the absorption along the line of sight through the relationship provided in \cite{Turatto2003} for "low reddening", obtaining a total absorption along the line of sight of $A_{V}$=0.8 $\pm$ 0.1 mag. At the same time, we tried to make use of the homogeneity observed for this class of objects during the plateau (\citealt{PastoLLSNe2004,SpiroPasto2014}): we estimated the absorption necessary to bring the colour evolution of SN 2021aai closest to the colour evolution of a sample of LL SNe IIP (taken from \citealt{2003Z,PastoLLSNe2004,Pasto2005cs}) between 30 and 100 days. Similar procedures were already performed, for example for SN 2001dc \citep{PastoLLSNe2004}. Through the method of the least squares, we obtained an absorption of $A_{V}$=1.92 $\pm$ 0.06 mag ($A_{V}$=2.09 $\pm$ 0.06 mag, accounting for the internal reddening of the Milky Way), which will be referred hereafter as "high reddening scenario". To compare the colour evolution of SN 2021aai with the LL SNe IIP colours available in the chosen sample, we converted the $r$ magnitudes (AB magnitudes system) into Johnson $R$ magnitudes (Vega magnitude system) by applying a constant correction measured through spectrophotometry (we adopt $r-R$ = 0.28 mag, the average value measured during the plateau phase).

The apparent light curves obtained during the six months of follow--up are shown in Figure \ref{SN2021aai_Phot}. 
The rise to maximum was not observed, but the explosion epoch was well constrained at MJD = 59223.4 $\pm$ 1.0, thanks to an upper limit ($r$ > 20.5 mag) obtained by ZTF just two days before the first detection. The plateau phase was unusually long--lasting, with a duration of 140 days: a tentative physical explanation will be discussed in Sect. \ref{Hydro}. During the plateau, the $r$ band displays a progressive brightening, spanning from $-15.87$ mag to --16.57 mag ($\pm$ 0.09 mag) in the high reddening scenario, and from --14.77 mag to --15.47 mag in the low reddening scenario. A similar behaviour is recorded in the NIR, where the transient became one magnitude brighter in the $J$, $H$ and $K$ bands from 13 d to 130 d.
The $g$ band evolution of SN 2021aai was different, with the transient reaching a peak magnitude of --16.41 (--14.84) mag at 5 days after the explosion, and then settling on a constant value of --15.68 (--14.11) mag up until the fall from the plateau in the high (low) reddening scenario.
During the fall from the plateau, which was well sampled in the $r$ and $i$ bands, there was a marked drop of 2.88 mag in 16 days.

\begin{figure}
\includegraphics[width=0.9\columnwidth]{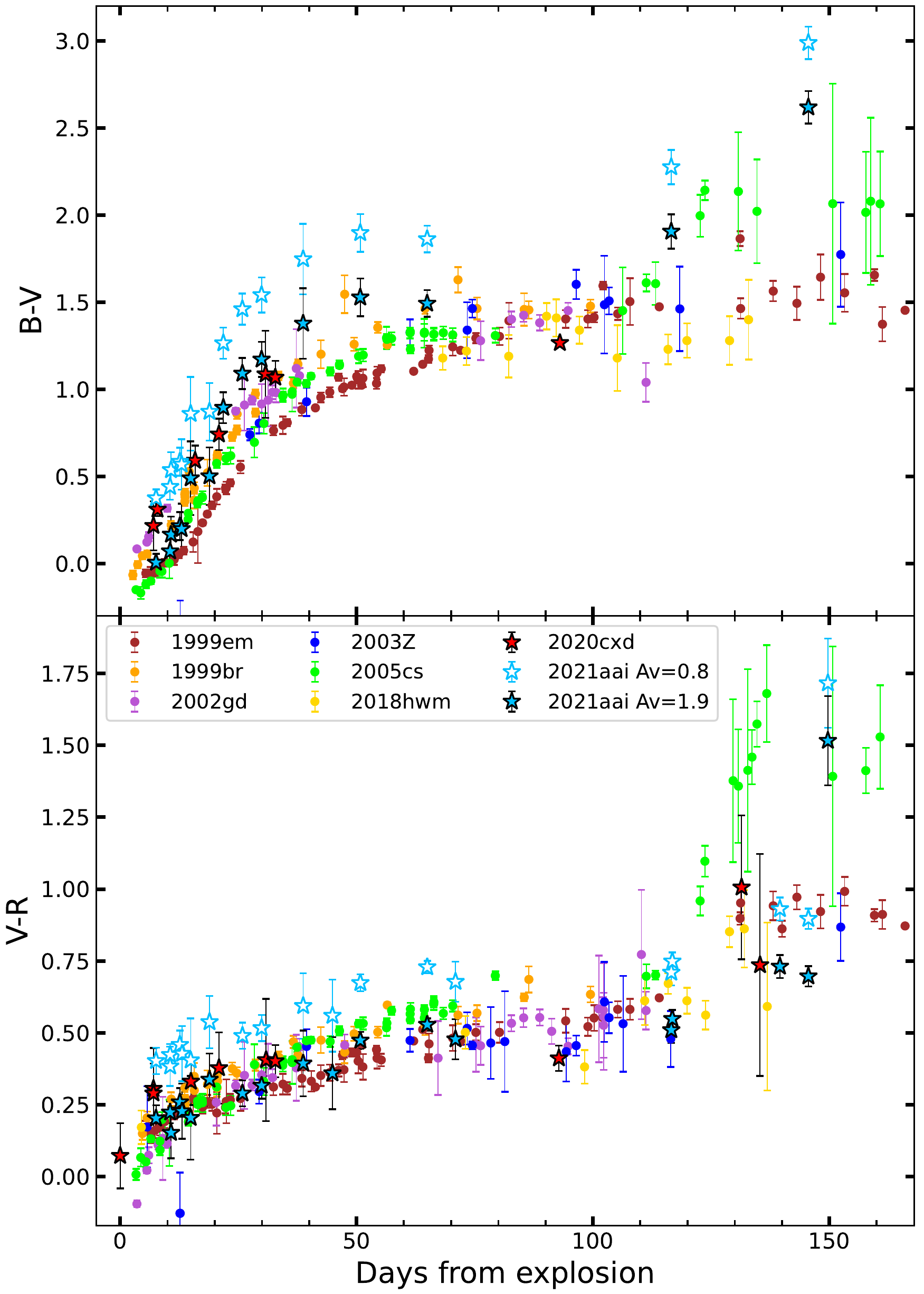}
\caption{$B-V$ and $V-R$ colour evolution for some of the objects presented in Figure \ref{Comparison}. SN 2021aai is reported twice, both with the low and the high reddening correction discussed in the text.}
\label{Colour}
\end{figure}


\begin{figure*}
\centering
\begin{minipage}[b]{.45\textwidth}
\includegraphics[width=1.0\columnwidth]{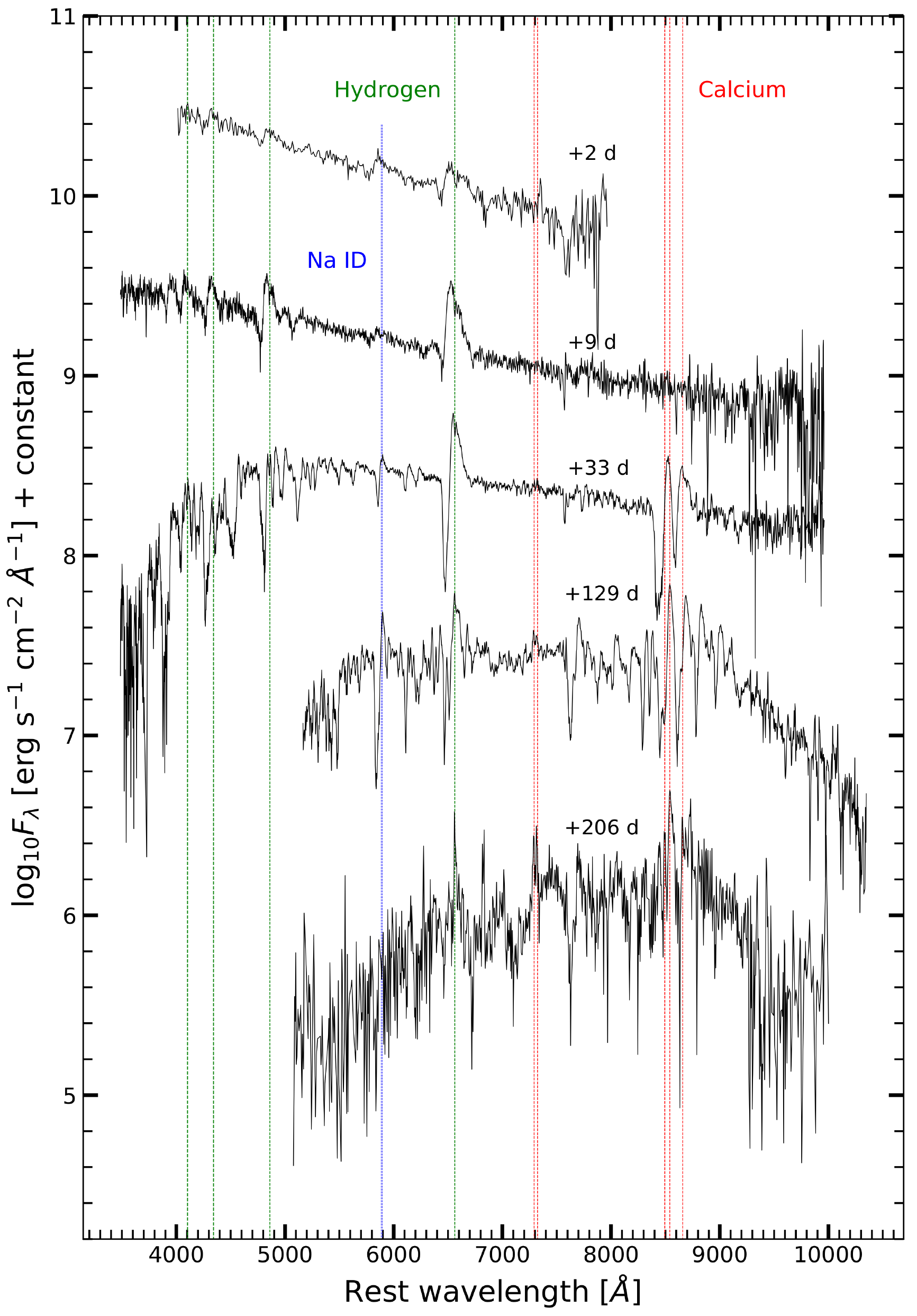}
\caption{Optical spectra of SN 2020cxd. Dashed lines mark the position of the Balmer series lines, Ca and Na I D lines. All spectra were corrected for reddening and redshift. Epochs are reported with respect to the explosion date.}
\label{SN2020cxd_Spec}
\end{minipage}\qquad
\begin{minipage}[b]{.45\textwidth}
\includegraphics[width=1.0\columnwidth]{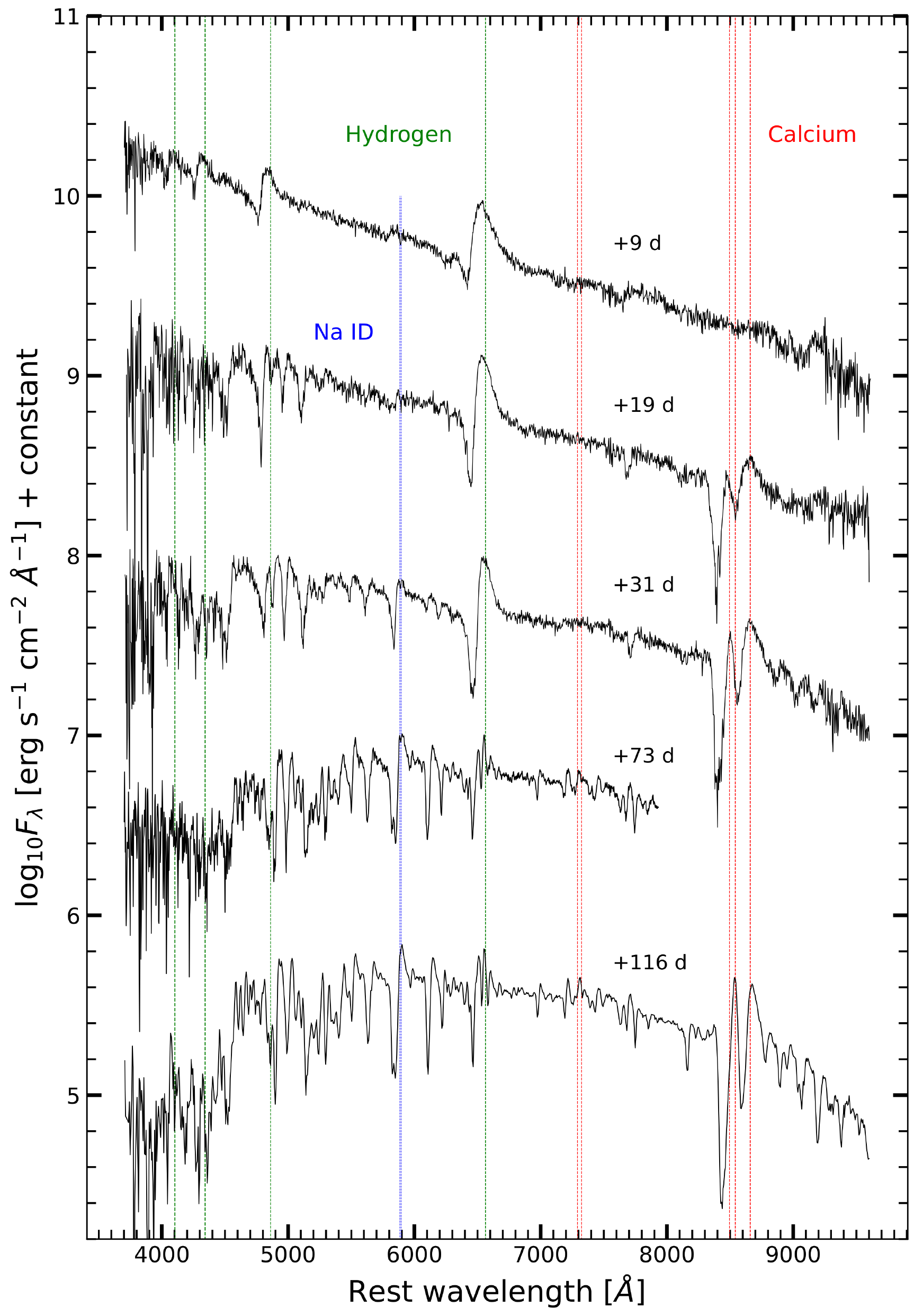}
\caption{Optical spectra of SN 2021aai. Dashed lines mark the position of the Balmer series lines, Ca and Na I D lines. All spectra were corrected for redshift and reddening according to the high reddening scenario. Epochs are reported with respect to the explosion date.}
\label{SN2021aai_Spec}
\end{minipage}
\end{figure*}

\subsection{Comparison with the LL SNe IIP class}
We compare SNe 2020cxd and 2021aai with LL SNe II and a borderline standard SN IIP that have good photometric and spectroscopic coverage. For this reason, we choose SN 1999br \citep{PastoLLSNe2004}, SN 1999em \citep{Phot1999em}, SN 2003Z \citep{SpiroPasto2014}, SN 2005cs \citep{Pasto2005cs}, SN 2010id \citep{2010id}, SN 2018hwm \citep{Reguitti2018hwm}.
In Figure \ref{Comparison}, we plot the $R$ band light curves for the chosen sample of faint SNe IIP. We convert the Sloan $r$ magnitudes of SN 2018hwm, SN 2020cxd and SN 2021aai to Johnson $R$ magnitudes by applying the constant correction discussed above for SN 2021aai ($r-R$ = 0.16 mag for SN 2020cxd, $r-R$ = 0.23 mag for SN 2018hwm). While relatively brighter objects like SN 2005cs or SN 2018hwm display a plateau at M$_{r}$ $\sim$ --15 mag, SN 2020cxd lies towards the low luminosity end of core collapse events, marked by the faint SN 1999br. SN 2021aai is located towards the brighter end of the peak luminosity distribution, especially in the high reddening scenario, when it is comparable to the standard event SN 1999em. The difference in the plateau luminosity can be physically interpreted as a different mass and density profile of the recombining H powering the light curve, a different expansion velocity of the ejected gas, or a different initial radius of the exploding star.
During the first 50 days of evolution, the light curve of SN 2020cxd closely resembles that of SN 2010id. However, the two light curves become different after $\sim$50 days, when SN 2020cxd shows a rebrightening while SN 2010id starts to fade. Indeed, both SNe 2020cxd and 2021aai are characterised by an increase of brightness towards the end of the plateau. This behaviour is not unheard of, as shown by \cite{GalbanySlope}, and it is more common in the red bands of faint transients with long plateau phases.
Indeed, the plateau of SN 2021aai is among the longest observed with its 140 days of duration, outlasting even the peculiar SN 2009ib \citep{Takatas2009ib}. For context, the average plateau duration for a SN IIP is 83.7 $\pm$ 16.7 days (obtained for the $V$ band by \citealt{AndersonAverageIIP}). SN 2021aai shows a late time decline close to that  band 2018hwm, while SN 2020cxd displays one of the faintest late time declines observed, even for LL SNe.

In Figure \ref{Colour}, we display the $B-V$ and $V-R$ colour evolution of SNe 2020cxd and 2021aai along with the colours observed for LL SNe IIP. Qualitatively, the behaviour of LL SNe IIP is quite homogeneous, as was already shown by \cite{SpiroPasto2014}. After a rapid increase in colour during the first 50 days ($\sim$ 1.5 mag increase in $B-V$ and $\sim$ 0.5 mag increase in $V-R$), the colours remain roughly constant up to $\sim$ 120 days, when SNe IIP typically fall from the plateau, leading to a final increase in colour as the transients become redder. The $g-r$ colour curve of SN 2020cxd, reported in the appendix (Figure \ref{Colour_appenidx}), shows an interesting behaviour after 120 days. We observe a steep increase in colour during the fall from the plateau, and a subsequent inversion in the trend as the colour $g-r$ becomes bluer. Such feature was pointed out for the first time for SNe 1997D and 1999eu \citep{PastoLLSNe2004}. As for SN 2021aai, it is possible to appreciate the difference in the colour evolution for the low and high reddening scenario, respectively. By construction, in the high reddening scenario the behaviour of SN 2021aai resembles more closely that of the other LL SNe IIP.



\begin{figure*}
\includegraphics[width=1.8\columnwidth]{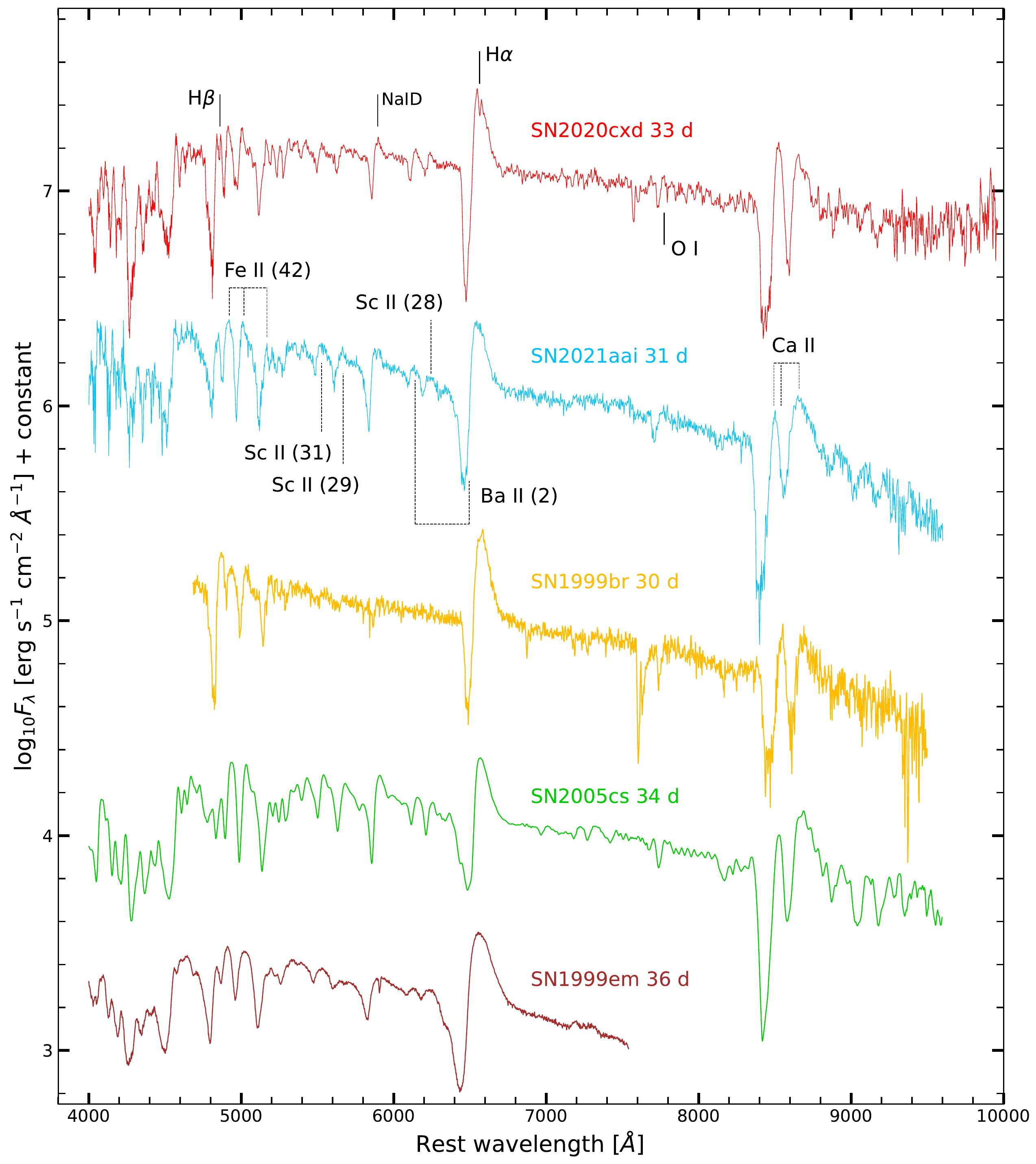}
\caption{Comparison of SN 2020cxd and SN 2021aai together with LL SNe IIP (SN 1999br and SN 2005cs) and a standard event (SN 1999em). All spectra were collected between 30 and 36 days after explosion and corrected for redshift and reddening (in the high reddening scenario for SN 2021aai). On SNe 2020cxd and 2021aai are marked the main spectral features characterising LL SNe IIP (at rest wavelength).}
\label{SpecComparison}
\end{figure*}

\section{Spectroscopic evolution} \label{Spectroscopic}

\begin{table}
	\centering
	\caption{Log of original spectroscopic observations for SN2020cxd and SN2021aai. Phases are reported with respect to the explosion epoch.}
	\label{LogSpectra}
	\begin{tabular}{llll} 
		\hline
Phase (days) & MJD &  Setup & Resolution [\AA]   \\
		\hline
		\hline
       	      &    \textbf{SN2020cxd}	   &     	&  \\
		\hline
2.3  &  58899.3    & LT+SPRAT & 18.0 \\
8.6  &  58905.6    & LCO+FLOYDS & 15.5 \\
32.5 &  58929.5    & LCO+FLOYDS & 15.5 \\
128.5 &  59025.5    & GTC+OSIRIS & 7.5 \\
205.7 &  59102.7   & GTC+OSIRIS & 8.0  \\
		\hline
		&    \textbf{SN2021aai}	   &     	&  \\
		\hline
8.5  &  59231.9    & NOT+ALFOSC & 14.6 \\
10.5  &  59233.9    & TNG+LRS & 15.5 \\
18.5 &  59241.9    & NOT+ALFOSC & 14.1 \\
30.5 &  59253.9    & NOT+ALFOSC & 14.1 \\
35.6 &  59259.0    & TNG+LRS & 15.6 \\
61.5 &  59284.9    & NOT+ALFOSC & 18.2 \\
72.5 & 59295.9    & TNG+LRS & 10.4 \\
115.5 &  59338.9    & NOT+ALFOSC & 14.1 \\
		\hline
	\end{tabular}
\end{table}

\subsection{Spectroscopic features}
Figures \ref{SN2020cxd_Spec} and \ref{SN2021aai_Spec} show the spectral sequences for SNe 2020cxd and 2021aai. The log of the spectroscopic observations is reported in Table \ref{LogSpectra}. In the first two spectra of SN 2020cxd, we notice a blue continuum: a blackbody fit yields a temperature of 10000 K at 2 d, which quickly declined to 8000 K at 9 d. Both spectra display prominent Balmer lines and few weaker lines, such as He I and Na ID displaying a P Cygni profile. The absence of the interstellar sodium absorption doublet leads us to estimate the internal absorption in the host galaxy as negligible.
At 30 d, we notice the arising of several new features: Ca II lines start to appear in the red part of the spectrum, in particular the forbidden doublet [CaII]  $\lambda\lambda$7291,7323 and Ca II NIR (Figure \ref{SN2020cxd_Spec}). On the blue part of the spectrum, several metal lines are identified, especially those of Fe II triplet 42 ($\lambda\lambda\lambda$4924,5018,5169), Sc II ($\lambda\lambda$5669,6246) and Ba II ($\lambda\lambda$6142,6497). Some of the most prominent metal lines are highlighted in Figure \ref{SpecComparison}, where it is also possible to appreciate the similarities between the spectra of SN 2020cxd and SN 2021aai.

The spectral features mentioned so far are extensively observed in LL SNe \citep{PastoLLSNe2004,SpiroPasto2014,ClaudiaGutierrez2020synthesis,Reguitti2018hwm}.
The presence of a relevant amount of metals gives rise to "line blanketing", where the flux in the bluest part of the spectrum is reduced by the metal absorption lines (see e.g. \citealt{MoriyaMazzali}). For this reason, when estimating the blackbody temperature from the continuum, it is important to exclude the blanketed region (indicatively, at wavelengths shorter than 5000 \AA) from the fit. Taking this effect into account, the blackbody fit of the continuum at 31 d yields a temperature of 5460 K, in line with the expectations for H recombination.
The last two spectra are taken after the drop from the plateau, during the late tail decline, when the [Ca II] doublet and Ca NIR triplet become prominent.

In Figure \ref{SN2021aai_Spec}, we present the spectral evolution of SN2021aai. We obtained a high quality sampling of the target during the plateau phase, but unfortunately it was impossible to follow the object after the fall from the plateau due to visibility constraints.
The first spectrum, at 8 d, is dominated by H lines. The interstellar Na I D absorption doublet is identified, suggesting a significant line of sight reddening towards SN 2021aai (see Sect. \ref{DiscoveryPhotometry}).
At later phases, the broad Na I D feature develops a clear P Cygni profile, at the same phases when the Ca II NIR triplet and the metal lines appear in the spectra. In Figure \ref{SpecComparison}, we compare the spectra at $\sim$ 30 d of SN 2020cxd and SN 2021aai with SN 1999br \citep{PastoLLSNe2004}, SN 2005cs \citep{Pasto2005cs} and SN 1999em \citep{Phot1999em}. The similarity among this sample of objects is striking, considering that they span over two magnitudes in peak luminosity. Beside the obvious P Cygni profile of H$\alpha$, all the objects are characterised by evident Ca II NIR triplet lines, Sc II $\lambda$6246 and Fe II multiplet 42 ($\lambda\lambda\lambda$4924,5018,5169). The differences lie, of course, in the line velocities: the position of the minimum of the P Cygni profile and the width of the H$\alpha$ feature in SN 1999em suggests a significantly higher expansion velocity for this object, which separates this standard SN IIP from the other LL SNe shown.

\begin{figure}
\includegraphics[width=1\columnwidth]{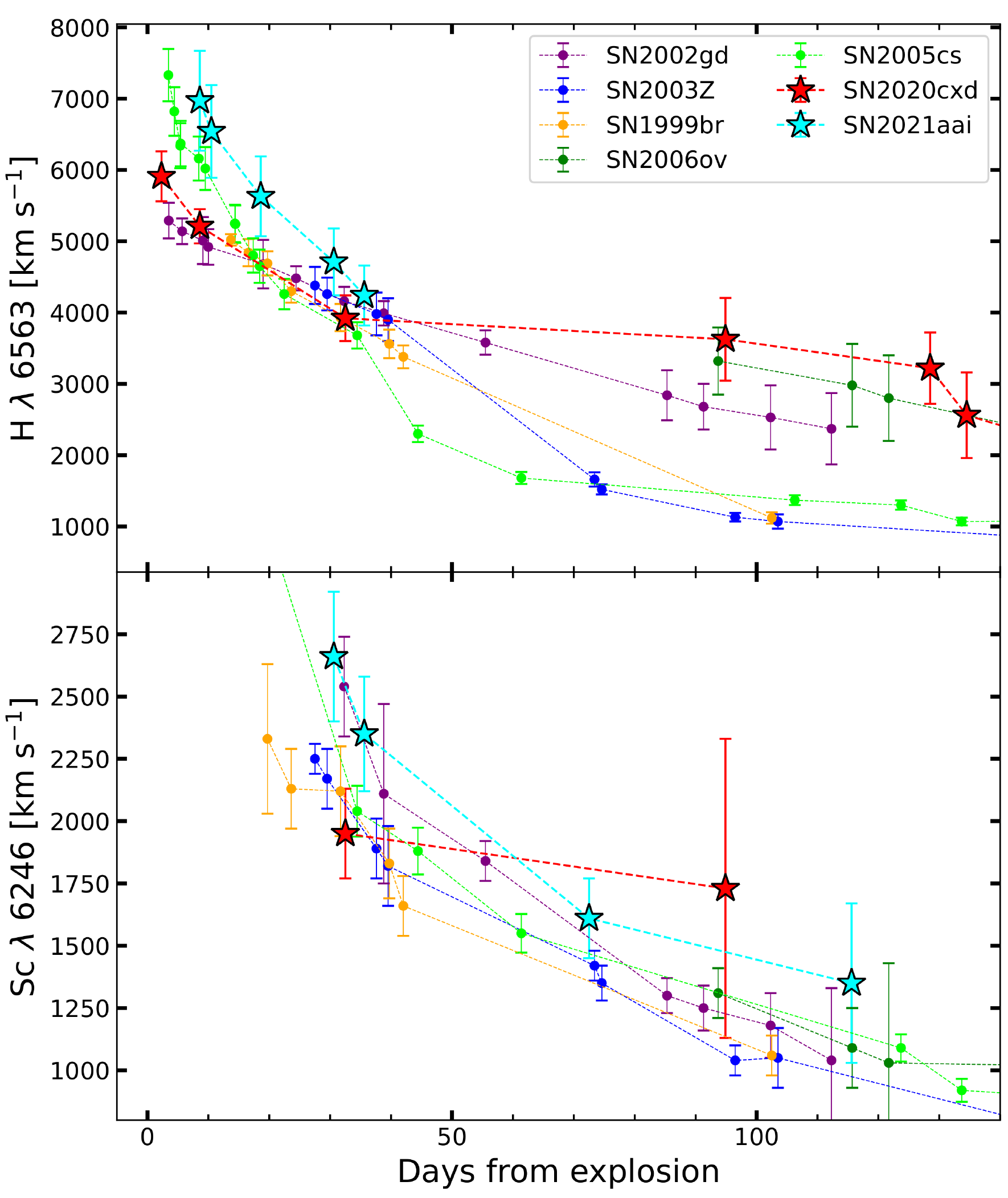}
\caption{Expansion velocities measured on the H$\alpha$ and Sc II $\lambda$ 6246 lines. The values obtained for SN 2020cxd and SN 2021aai are compared with those of other LL SNe IIP. }
\label{LineVel}
\end{figure}

\subsection{Expansion velocities}

We estimate the velocity of the expanding gas by measuring the position of the minimum of the P Cygni absorption profiles. Different species yield a different expansion velocity, reflecting a different position where the line forms through the ejecta \citep{Gutierrez17a}. Due to higher optical depth, H$\alpha$ and H$\beta$ lines form in the outer layers of the expanding materials, therefore yielding higher velocities than other species. Fe II lines, especially those belonging to multiplet 42, have a lower optical depth, and have been widely used to estimate the expansion velocity of the ejecta at the photosphere \citep{Hamuy2003}. The Sc II line $\lambda$6246 displays an even lower optical depth, and is sometimes used as a proxy for expansion velocity instead of the Fe II lines (e.g. \citealt{Maguire2010}). 
For SN 2020cxd in particular, the velocity measurements performed on the H$\alpha$ line showed that the line forming region moves in velocity space monotonically from 5900 km s$^{-1}$ at 2 d, to 2560 km s$^{-1}$ immediately after the drop from the plateau (134 d), and finally to 1020 km s$^{-1}$ at 245 d.
The H$\alpha$ expansion velocity after 90 days is measured from the Full Width at Half Maximum (FWHM) of the line, since the rise of the Ba II $\lambda$ 6497 makes it impossible to identify clearly the position of the minimum of the P Cygni profile. The results are reported in Table \ref{TableLines} and plotted in Figure \ref{LineVel}, along with other values from LL SNe IIP taken from \cite{PastoLLSNe2004,Pasto2005cs} and \cite{SpiroPasto2014}. From the comparison with similar objects, we notice that SN 2020cxd displays low H$\alpha$ and Sc II expansion velocities in the early phases (before 50 d), compatible with the values obtained for SN 1999br \citep{PastoLLSNe2004}. Later epoch values, however, appear to be more in line with higher velocity objects like SN 2006ov \citep{SpiroPasto2014}. It is important to notice, especially for the Sc II measurements at 95 d, that the resolution of the spectrum was poor, leading to a large error.

For SN 2021aai, the velocities measured from the H$\alpha$ P Cygni profiles range from 7000 km s$^{-1}$ at 8 d to 4200 km s$^{-1}$ at 35 d. Subsequently, the rise of the Ba II $\lambda$ 6497 line in the absorption part of the P Cygni profile forces us to estimate the expansion velocities from the FWHM of the emission component of the H$\alpha$ line, as previously done by \cite{Sheng2020cxd} for SN 2020cxd. As already mentioned, metal lines are characterised by a lower optical depth, leading to their formation closer to the photosphere compared to H lines, which form in the outer layers of the ejecta and therefore yield higher velocity measurements. Both H$\alpha$ and Sc II expansion velocities for SN 2021aai are shown in Figure \ref{LineVel}. SN 2021aai shows high velocities both in the H I and Sc II measurements, located consistently at the top end of the velocity distribution for the sample of objects considered. Since it is also among the most luminous LL SNe (adopting the high reddening scenario), this would favour the interpretation in which SNe IIP are characterised by a continuum of properties, spanning from LL SNe IIP to the most luminous ones, with brighter objects showing higher velocities and a larger ejected \textsuperscript{56}Ni mass, as suggested by \cite{PastoLLSNe2004}. Such correlation will be discussed in more detail in Sect. \ref{NiEst}.

\begin{table}
	\centering
	\caption{Expansion velocities measured for relevant lines through the position of the minimum of the P Cygni absorption profile. All velocities are in km s$^{-1}$. Measurements for SN 2020cxd taken at 94.9 and 134.5 d were performed on spectra presented in \protect\cite{Sheng2020cxd}. }
	\label{TableLines}
	\begin{tabular}{lllll} 
		\hline
Phase (days) & Sc II $\lambda$ 6246 & Fe II $\lambda$ 5169 & H$\beta$ & H$\alpha$ \\
		\hline
		\hline
     &  	      &    \textbf{SN 2020cxd}	   &     	&  \\
		\hline
2.3      &       --    &     --     &  5670  (430) &  5910  (350) \\
8.6      &       --    &      4800  (600) & 5240  (400) &  5210  (240) \\
32.5     &    1950  (180)  &  3020  (250) & 3520  (300) &  3920  (320) \\
94.9     &    1730  (600)  &  --    &   --      &  3625  (580) \\
128.5     &    --  &  --    &   --      &  3220  (500) \\
134.5     &   --   &  --    &   --      &  2560  (600) \\
		\hline
     &  	      &    \textbf{SN 2021aai}	   &     	&  \\
		\hline
8.5     &  	    --  &    --	   &     6480 (970)	& 6970 (700) \\
10.5    &   --   &   5180 (620) &	    6170 (930) &	 6540 (650) \\
18.5    &	  --   &   3850 (480)	&    4810 (720)	& 5630 (560) \\
30.5    & 2660 (400) & 3020 (420)	  &  3580 (540)	& 4710 (470) \\
35.6    &	 2350 (350) & 2500 (380)	 &   2840 (430)	& 4240 (420) \\
72.5  	&	    1610 (240) & 1970 (340)	 &   --	  &    -- \\
115.5   & 	    1350 (320) & --	    &    --	  &    -- \\
		\hline
	\end{tabular}
\end{table}

\begin{figure}
\includegraphics[width=1.0\columnwidth]{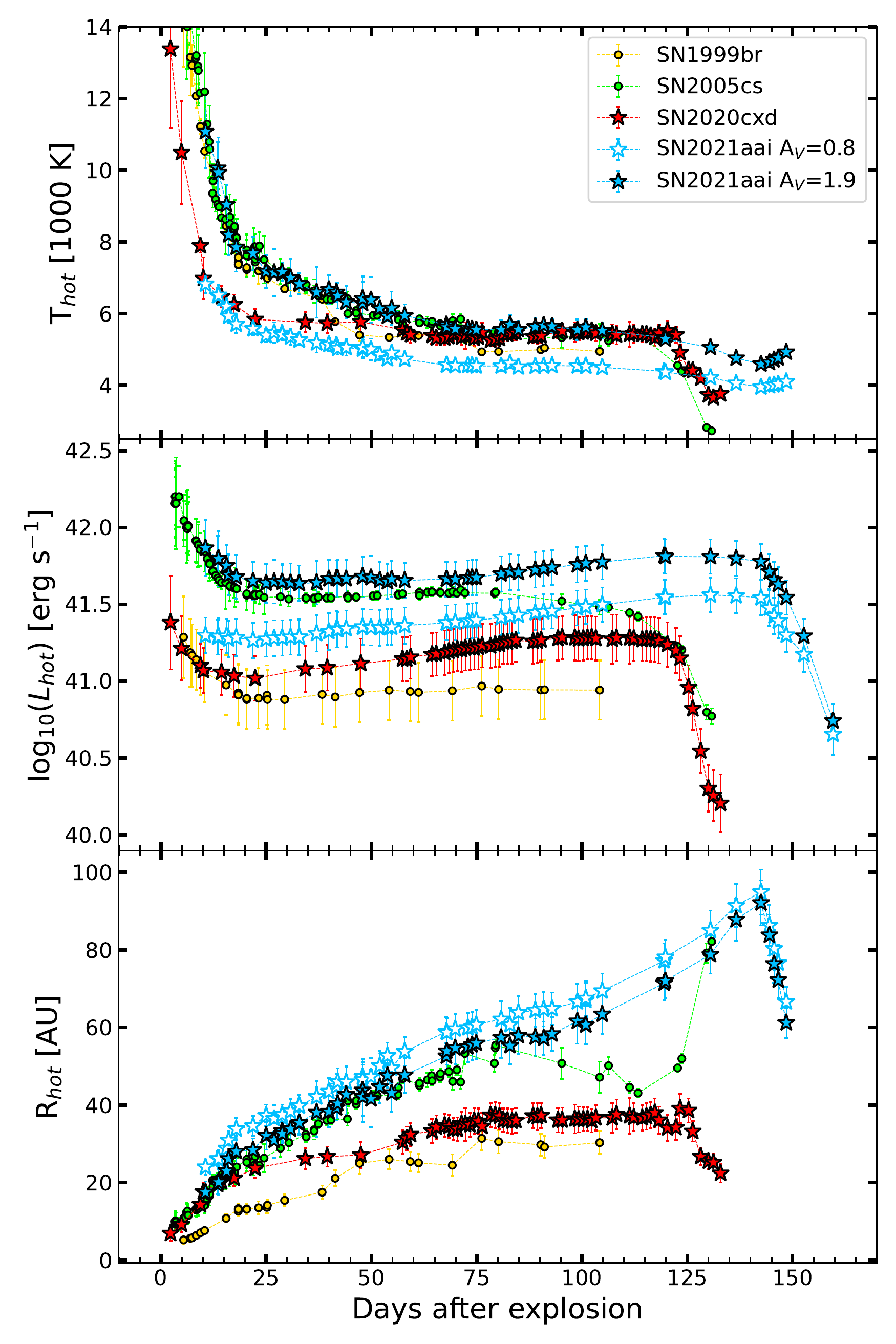}
\caption{Temperature, luminosity and radius evolution of SNe 2020cxd and 2021aai, along with SN 1999br and SN 2005cs for comparison. See text for the details about the blackbody fitting procedure. }
\label{TLR}
\end{figure}

\section{Blackbody Fitting} \label{BlackBodyF}
In order to estimate physical parameters characterizing SNe 2020cxd and 2021aai, we perform blackbody fits both on our photometric data and on our spectra. For the spectra, we use the \texttt{nfit1d} task in the \texttt{IRAF} package \texttt{stsdas}, fitting the continuum with a blackbody function.
For the fit of the photometric points, we perform a Monte Carlo simulation for each epoch, fitting with the \texttt{python} tool \texttt{curve\_fit}\footnote{https://docs.scipy.org/doc/scipy/reference/generated/ scipy.optimize.curve\_fit.html} 200 sets of fluxes randomly generated with a Gaussian distribution centered on the measured value, and $\sigma$ equal to the measured error. Such procedure is described in detail in \cite{LRNeHugsI}. Both in the spectroscopic and photometric fits, we exclude the regions heavily affected by line blanketing, since they would misleadingly reduce the estimated temperature. After obtaining a blackbody fit to the SED of the target (which already yields the temperature), we integrate it over wavelength and obtain the total flux emitted. Adopting the distances given in Sect. \ref{DiscoveryPhotometry} and assuming spherical symmetry, we calculate the bolometric luminosity of the source. Finally, the radius is estimated through the Stefan--Boltzmann law.
The temperature, luminosity and radius obtained for SNe 2020cxd and 2021aai are presented in Figure \ref{TLR}, together with the same values obtained for SN 1999br \citep{PastoLLSNe2004} and SN 2005cs \citep{Pasto2005cs2006}. On the top panel we see that SN 2020cxd displayed a very hot continuum (> 13000 K) at 2 d, quickly declining over the following days. At 22 days, the temperature already settles at $\sim$ 5500 K, corroborating the results obtained in Sect. \ref{Spectroscopic}. At 121 days, the temperature starts declining, along with the luminosity, as the object fades from the plateau.
The bolometric luminosity of SN 2020cxd is presented in the middle panel of Figure \ref{TLR} and shows a clear dip from 2.4 $\times$ 10$^{41}$ erg s$^{-1}$ at 2 d to 1.0 $\times$ 10$^{41}$ erg s$^{-1}$ at 22 d. During the following 90 days the transient steadily rebrightens, reaching 1.9 $\times$ 10$^{41}$ erg s$^{-1}$ at 111 d, before finally falling from the plateau at 120 d.
The radius (bottom panel) of the emitting blackbody quickly rises from 7 to 26 AU in the first 30 days, followed by a slower increase. Between 50 and 120 days, the emitting radius remaines roughly constant at $\sim$35 AU. When SN 2020cxd is fading from the plateau, the radius shows a decrease, which can be interpreted as the photosphere receding before the ejecta finally becomes transparent.
We do not fit a blackbody to the epochs in the linear decline, as the transient is transitioning from the photospheric to the nebular phase, where the luminosity is mostly supported by lines rather than continuum opacity.

For SN 2021aai, we discuss both the low reddening case with $A_{V}$=0.8 mag, obtained through the Na I D doublet absorption EW, and the high reddening case with $A_{V}$=1.9 mag, obtained through the colour comparison with other LL SNe IIP. The low reddening scenario is characterized by lower temperatures at all epochs, with a plateau temperature of only 4300 K. The high reddening scenario is much more promising in this situation, since the plateau temperature of SN 2021aai overlaps with the rest of the sample, at around 5500K. In particular, SN 2021aai in the high reddening case displays the same temperature evolution as SN 2005cs, and it is only marginally brighter when considering the bolometric luminosity. The clearest difference between the two objects is the duration of the plateau: for SN 2005cs the plateau ends  $\sim$120 days after the explosion, but the luminosity starts fading by $\sim$ 75 days. SN 2021aai, on the other hand, is definitely longer--lasting. In the high reddening scenario, its bolometric luminosity has an early peak (7.2 $\times$ 10$^{41}$ erg s$^{-1}$), similar to the other LL SNe IIP considered. After few weeks of dimming, SN 2021aai luminosity increases from 4.3 $\times$ 10$^{41}$ erg s$^{-1}$ at 25 d to 6.5 $\times$ 10$^{41}$ erg s$^{-1}$ at 130 d before the fall from its plateau. On the other hand, in the low reddening scenario there is no evidence of the early luminosity peak, and the bolometric luminosity steadily increases from 1.9 $\times$ 10$^{41}$ erg s$^{-1}$ to 3.6 $\times$ 10$^{41}$ erg s$^{-1}$ during the plateau phase.
 Unfortunately, we do not have enough multi--band observations or spectra during the first 10 days to perform a blackbody fit to confirm if the similarity between SN 2021aai (in the high reddening scenario) and SN 2005cs is present at the earliest phases. 
The larger luminosity of SN 2021aai compared to SN 2020cxd leads to an estimate of a larger radius, given that their plateau temperature was comparable. While starting off with similar values, the emitting radius of SN 2021aai grows much more than the one of SN 2020cxd, up to 95 AU at 143 days after the explosion. This behaviour appears to be unusual, compared to the other objects, where the radius varies significantly less during the plateau phase.

\begin{figure}
\includegraphics[width=1\columnwidth]{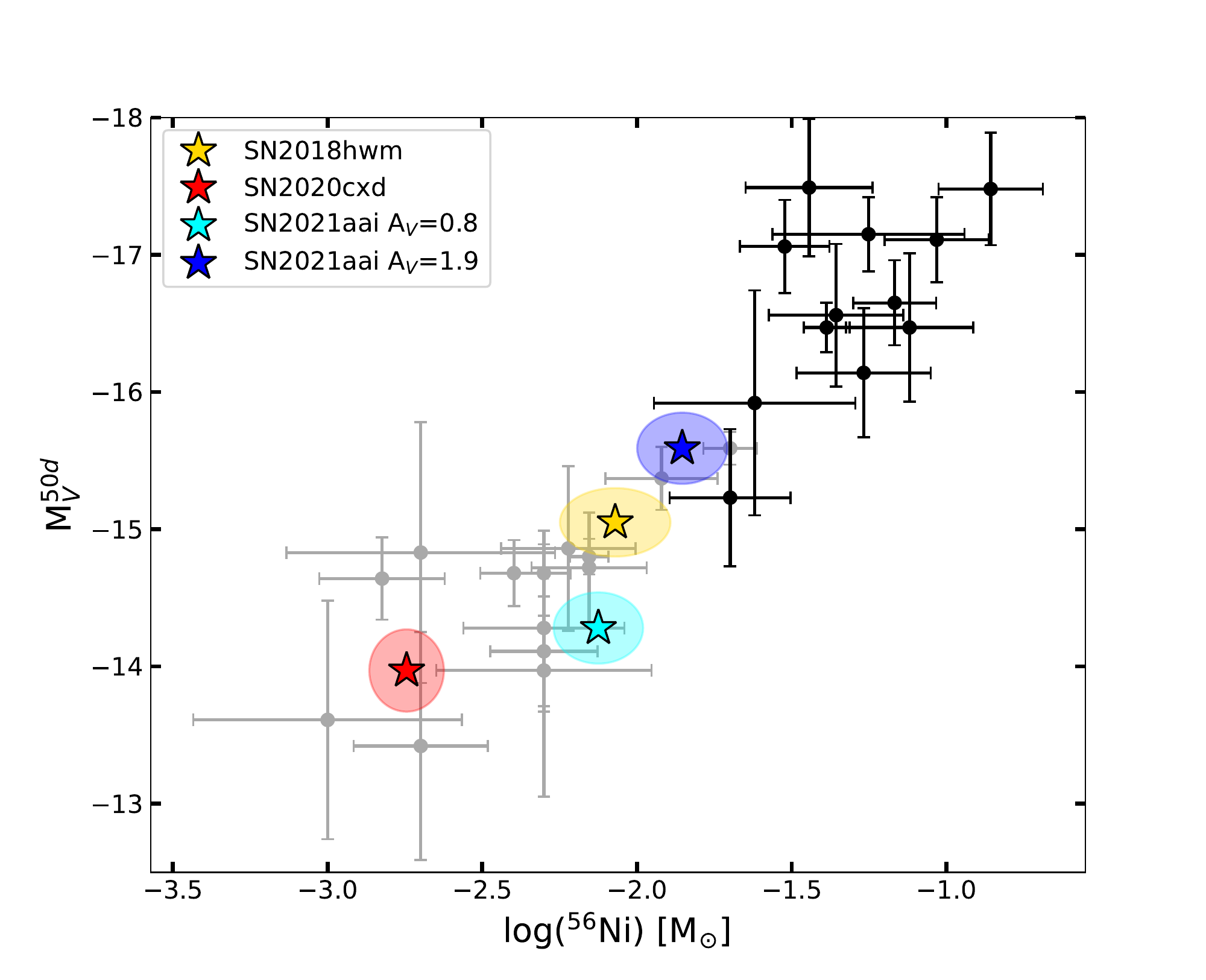}
\label{Ni56}

\includegraphics[width=1\columnwidth]{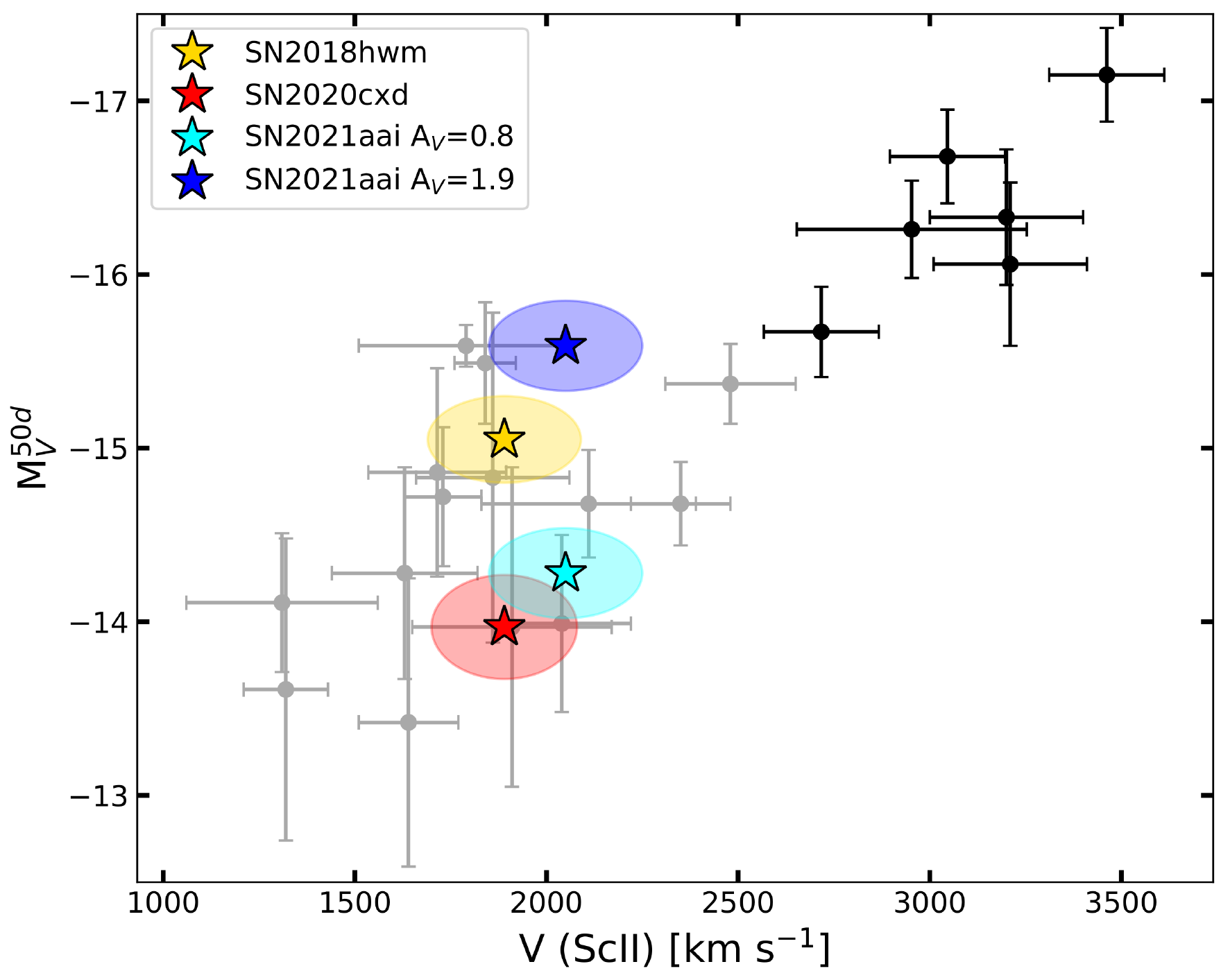}
\caption{Top panel: $V$ band absolute magnitude at 50 days versus \textsuperscript{56}Ni ejected mass. LL SNe are shown in grey \citep{PastoLLSNe2004,SpiroPasto2014,JagerPoint}, while standard SNe IIP are shown in black \citep{Rodriguez}. Some relevant objects are reported as coloured stars, with their errors shown as elliptical regions. Lower panel: same as top panel, but with expansion velocity of Sc II $\lambda$6246 \citep{Maguire2010} instead of \textsuperscript{56}Ni mass.}
\label{ConfNiMag}
\end{figure}

\section{\textsuperscript{56}Ni Estimate} \label{NiEst}
The late tail of the light curve of SNe IIP is powered by the \textsuperscript{56}Ni $\rightarrow$ \textsuperscript{56}Co $\rightarrow$ \textsuperscript{56}Fe decay chain, which deposits energy into the expanding gas in the form of photons and positrons \citep{ColgateMcKee1969}. We estimate the ejected mass of \textsuperscript{56}Ni through a comparison of the late time luminosity with the well studied SN 1987A, as previously done for other LL SNe IIP (e.g. \citealt{PastoLLSNe2004,SpiroPasto2014,Tomasella2018}), through the following equation:

\begin{equation}
    M(\textsuperscript{56}Ni)_{SN} = M(\textsuperscript{56}Ni)_{1987A} \times \frac{L_{SN}}{L_{1987A}}
    \label{Nickel}
\end{equation}

where we adopt a value for the \textsuperscript{56}Ni ejected mass by SN 1987A of 0.073 $\pm$ 0.012  M$_{\odot}$, which is the weighted average of the values reported in \cite{ArnetNi1} and \cite{BouchetNi2}.
Due to a lack of information in the NIR during the late decline, we have to perform some approximations. We compare the integrated luminosity in the observed bands ($r,i,z$) for our objects with the luminosity integrated through the same wavelength ranges for SN 1987A (since SDSS filters were not available at the time).
With this method, we obtain for SN 2020cxd (1.8 $\pm$ 0.5) $\times$ 10$^{-3}$ M$_{\odot}$ of synthesised \textsuperscript{56}Ni, quite low compared to the typical value of few 10$^{-2}$ M$_{\odot}$ for a SN IIP event (see, for example $M$(\textsuperscript{56}Ni)$_{avg}$ = 0.033 $\pm$ 0.024 M$_{\odot}$ obtained by \citealt{AndersonAverageIIP}).
For SN 2021aai, we obtain a value of (7.5 $\pm$ 2.5) $\times$ 10$^{-3}$ M$_{\odot}$ for the low reddening scenario and (1.4 $\pm$ 0.5) $\times$ 10$^{-2}$ M$_{\odot}$ for the high reddening scenario, which is still a factor of 2 below the average SN IIP event reported by \cite{AndersonAverageIIP}.

\begin{figure*}

\includegraphics[width=1.5\columnwidth]{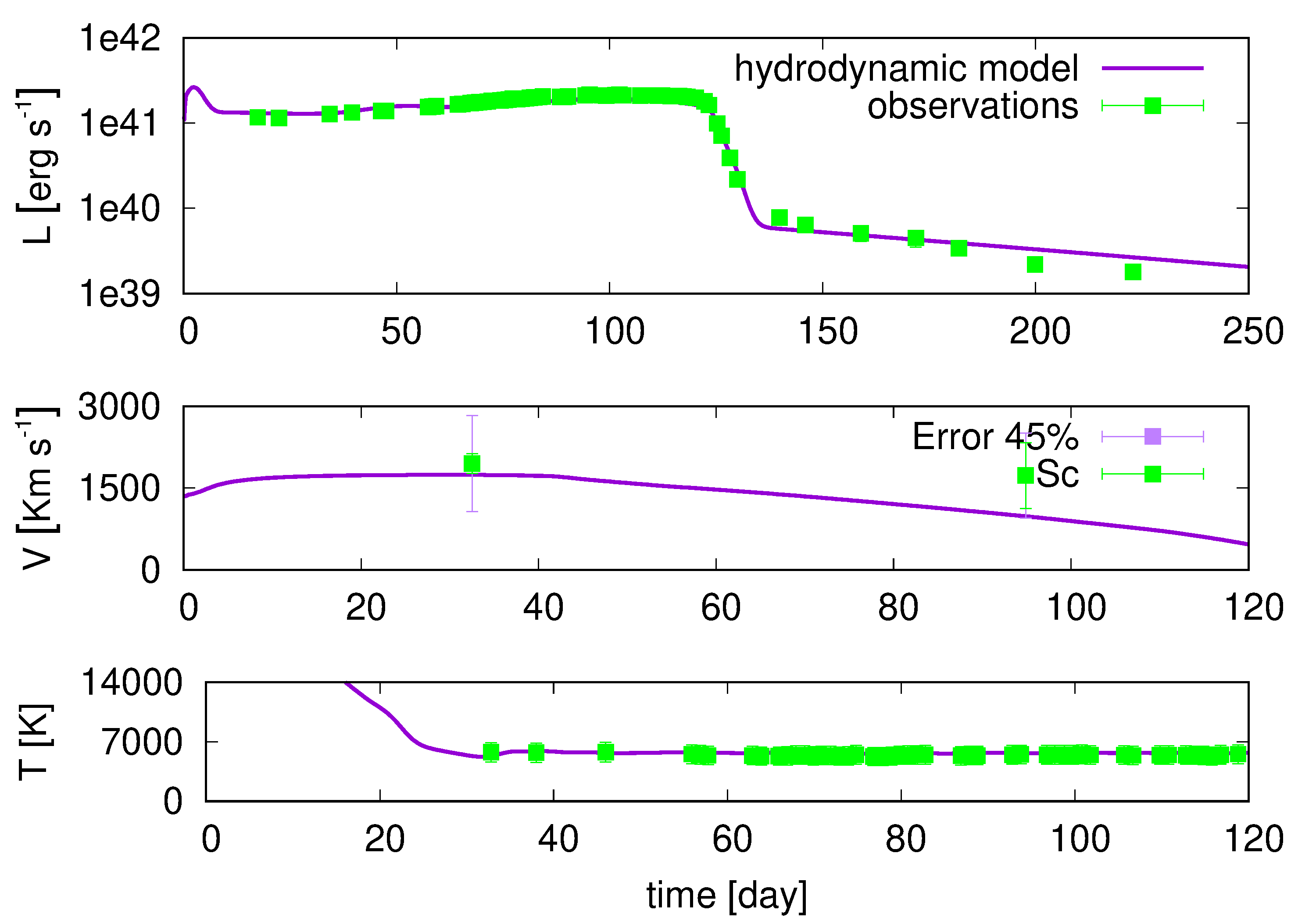}
\caption{Evolution of the main observables of SN 2020cxd compared to the best hydrodynamical model. The parameters characterizing the displayed fit are $R$ = 4 $\times$ 10$^{13}$ cm ($\sim$ 575 R$_{\odot}$), $M_{ej}$ = 7.5 M$_{\odot}$, and $E$ = 0.097 foe (see text for details). In the top panel, the bolometric luminosity is displayed. In the middle panel, the photospheric velocity obtained through the ScII lines as described in Sect. \ref{Spectroscopic}. Notice that the second velocity measurement is affected by a large error due to poor spectral resolution, as displayed in Figure \ref{LineVel}. Finally, in the bottom panel is shown the temperature evolution.}
\label{HydroCXD}
\end{figure*}

\begin{figure*}

\includegraphics[width=1.5\columnwidth]{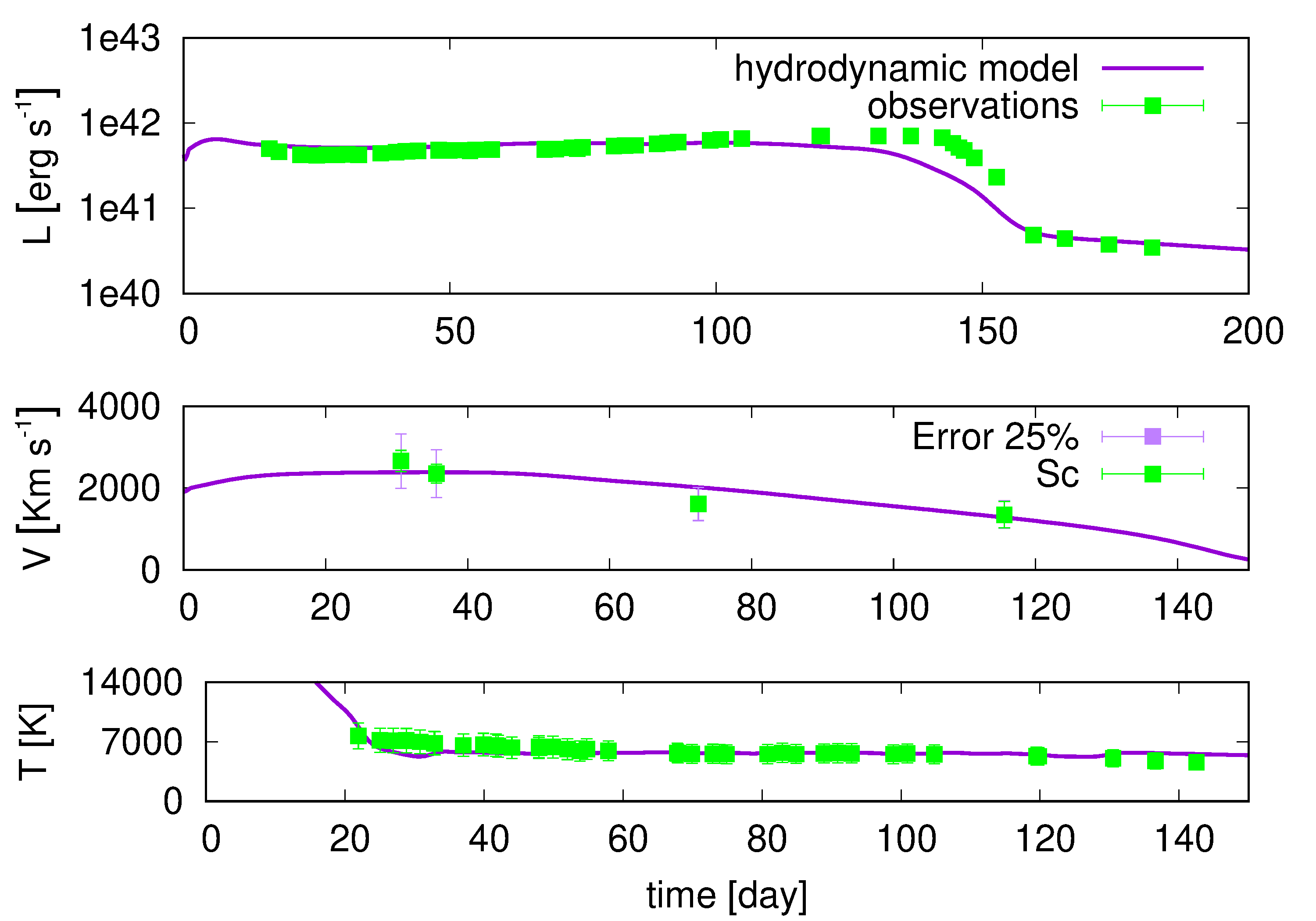}
\caption{Same as Figure\ref{HydroCXD}, but for SN 2021aai in the high reddening scenario. The parameters characterizing the displayed fit are R = 4 $\times$ 10$^{13}$ cm ($\sim$ 575 R$_{\odot}$), M$_{ej}$ = 15.5 M$_{\odot}$, and E = 0.4 foe (see text for details) In this case, the observed ScII lines velocities are more reliable, and better reproduced by the model. At the same time, the bolometric luminosity shows a more extended plateau compared to our fit.}
\label{HydroAAI}
\end{figure*}

In Figure \ref{ConfNiMag}, we display the locations of SN 2020cxd and SN 2021aai in the peak magnitude--\textsuperscript{56}Ni ejected mass plane and the peak magnitude versus expansion velocity plane for SNe IIP, both introduced in \cite{Hamuy2003}. As we can see there is no clear separation between standard and LL SNe IIP, but rather a smooth transition between the two classes. According to the classification adopted in the literature, SN 2021aai in the high reddening scenario would be in the transition region between low luminosity and standard objects, when considering the \textsuperscript{56}Ni ejected mass. Instead, SN 2020cxd is definitely in the lowest end of the parameter spectrum. Considering the expansion velocities measured with Sc II $\lambda$6246, instead, both objects display average values for LL SNe IIP.

\section{Hydrodynamical Modelling} \label{Hydro}
\subsection{Model details}
In order to estimate the physical properties of SNe 2020cxd and 2021aai at the explosion time (progenitor radius R, explosion energy E, total ejected mass M$_{ej}$), we use the hydrodynamical modelling procedure presented in detail in \cite{PumoModel2017}, and already well--tested on both faint and standard SNe IIP (e.g. \citealt{SpiroPasto2014,2009N,Tomasella2018,Reguitti2018hwm}).
The procedure consists in a simultaneous $\chi^{2}$ minimisation aiming at reproducing the observed bolometric luminosity, expansion velocity and photospheric temperature. This operation is performed in two distinct steps.
Firstly, a preliminary investigation is carried out through the model presented by \cite{ZampPrel2003}, solving the energy balance equation under the assumptions of ejecta with constant density in homologous expansion. The parameters obtained during this first fit lay down the framework on which the subsequent detailed calculations are based.
The second step makes use of a general--relativistic, radiation--hydrodynamics Lagrangian code \citep{PumoZampieriTuratto2010,PumoEZampieriModel}, which reproduces the main observables of the SN, from the onset of the plateau phase up to the nebular phase. The code takes into account the gravitational effects of the compact remnant left by the core collapse and the energy input from the decay of radioactive isotopes synthesised during the explosion.
It is important to note that we did not try to reproduce the early phase of the explosions ($\sim$ 15--20 days after explosion), since temperature and luminosity during this phase are significantly affected by emission from the outermost shell of the ejecta, which is not in homologous expansion, rendering the assumptions in our model inaccurate.
The best fitting models for SNe 2020cxd and 2021aai are shown in Figures \ref{HydroCXD} and \ref{HydroAAI} respectively.

\subsection{SN 2020cxd results and progenitor scenarios} \label{cxdECSN}
Adopting the \textsuperscript{56}Ni masses inferred in Sect. \ref{NiEst} and the well constrained explosion epochs in Sect. \ref{DiscoveryPhotometry}, we find the initial parameters of the progenitor of SN 2020cxd to be: $R$ = 4 $\times$ 10$^{13}$ cm ($\sim$ 575 R$_{\odot}$), M$_{ej}$ = 7.5 M$_{\odot}$, and E = 0.097 foe (sum of kinetic and thermal energy). The errors on the free model parameters reported due to the $\chi^{2}$ fitting procedure are about 15\% for M$_{ej}$ and R, and 30\% for E. 
To obtain the main sequence (MS) mass of the progenitor star of SN 2020cxd, we need to account for the compact remnant produced by the core collapse (1.3 -- 2.0 M$_{\odot}$) as well as the mass lost during the pre--SN evolutionary phases ($\lesssim$ 0.1 -- 0.9 M$_{\odot}$, as prescribed in \citealt{PumoModel2017}). Considering these corrections, the MS mass of the progenitor of SN 2020cxd is estimated to be 8.9 -- 10.4 M$_{\odot}$. 
We note that, despite the different methodology applied, our results are consistent with those obtained by \cite{Kozyreva2022_cxd}: M$_{ej}$ = 7.4 M$_{\odot}$, E = 0.07 foe and R = 408 R$_{\odot}$.

\begin{figure}
\includegraphics[width=1\columnwidth]{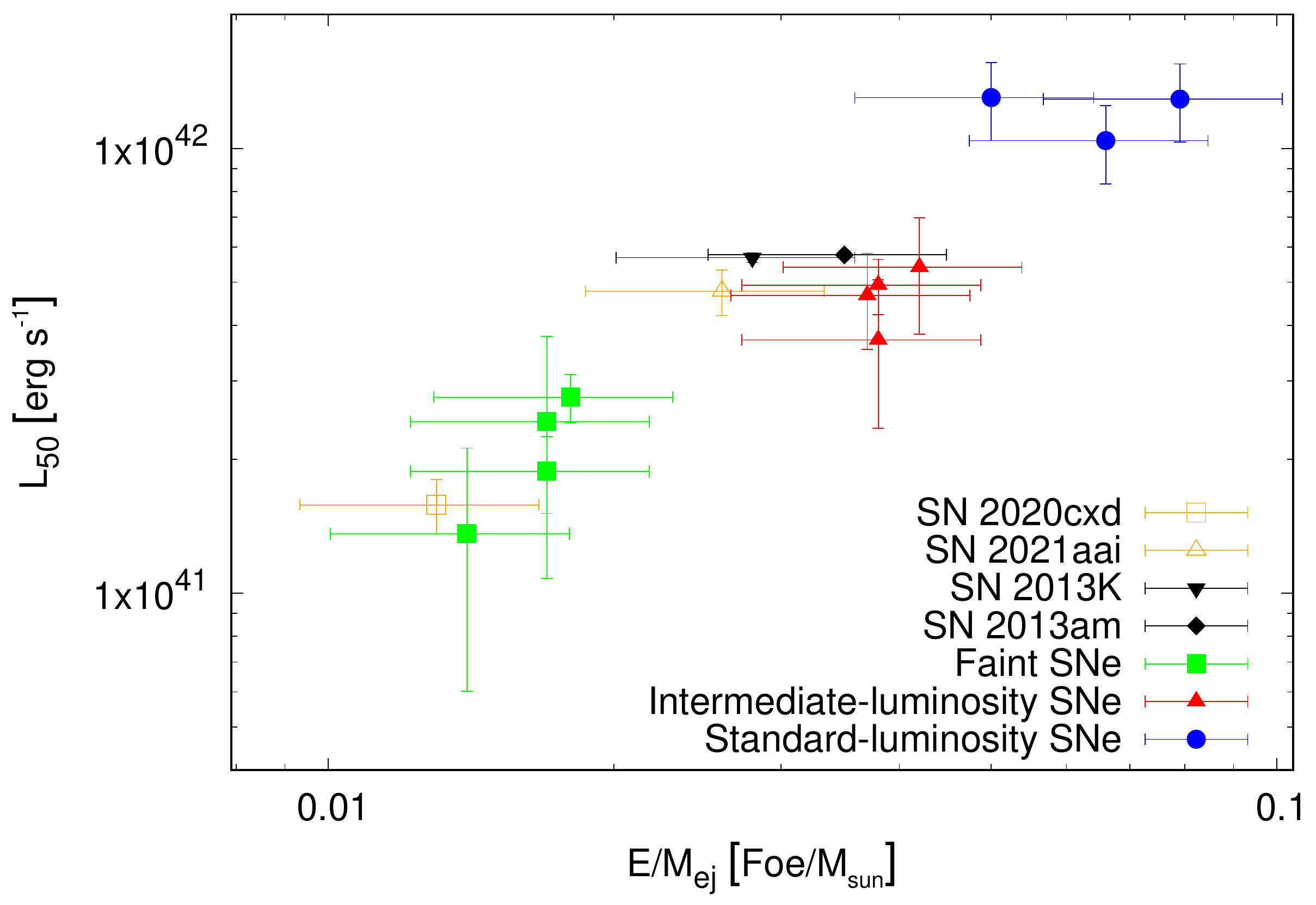}

\includegraphics[width=1\columnwidth]{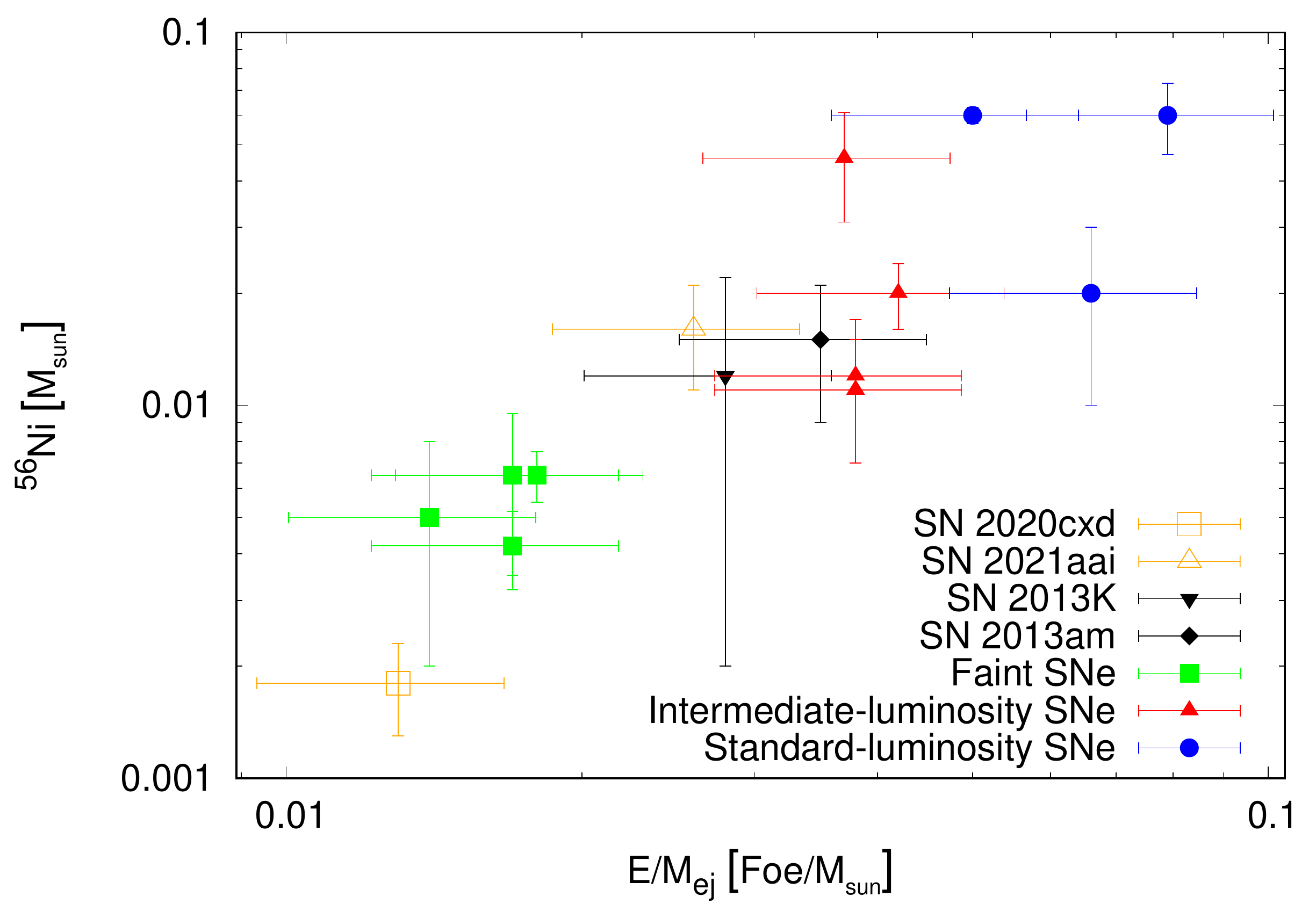}
\caption{Correlations between the plateau luminosity (top panel) and \textsuperscript{56}Ni mass (bot panel)  with the E/M$_{ej}$ ratio. LL SNe are coloured in green, standard SN IIP are shown in blue, while transitional objects are displayed in red and black \citep{PumoModel2017,Tomasella2018}. SNe 2020cxd and 2021aai are marked with orange symbols.}
\label{E/M2}
\end{figure}

The parameters estimated through hydrodynamical modelling are compatible with what is expected for a red supergiant (RSG) star. The radius is within the 500--1500 R$_{\odot}$ range associated with RSG, although leaning towards the lower end of the distribution, as reported in the review of \cite{SmarttProg}. Furthermore, the progenitor initial mass is just above the 8 $\pm$ 1 M$_{\odot}$ threshold that defines the minimum progenitor mass needed to produce a SN explosion, based on direct detections of RSG progenitors of SNe IIP \citep{SmarttProg}. For these reasons, SN 2020cxd could be explained by the explosion of a low mass RSG, resulting in the emission of a limited amount of energy compared to the explosion of more massive RSG. This corroborates the scenario where more massive RSG explode in SNe that are brighter and with faster ejecta compared to the explosion of less massive RSGs, which most likely produce LLSNe IIP \citep{PastoLLSNe2004,Tomasella2018}.
In this context, we display in Figure \ref{E/M2} the correlation between the plateau luminosity and \textsuperscript{56}Ni with the parameter E/M$_{ej}$, as in \citealt{PumoModel2017} (see their table 2, figs 5 and 6), including also the two ``intermediate-luminosity'' objects presented in \citealt{Tomasella2018} (i.e. SNe 2013K and 2013am). Like in \citealt{PumoModel2017} (to which we refers for details), the error bars on the E/M$_{ej}$ ratios are estimated by propagating the uncertainties on $E$ and M$_{ej}$, adopting a value of 30\% for the relative errors of E and 15\% for that of M$_{ej}$. Both in the top and bottom panel of Figure \ref{E/M2}, SN 2020cxd is at the very end of the distribution of SN IIP, due to the low E/M$_{ej}$ ratio inferred for the explosion and the relatively low amount of \textsuperscript{56}Ni synthesised.

Considering its faint nature and the inferred best--fitting model parameters, SN 2020cxd also appears to be a fair candidate for being an ECSN from a super-asymptotic giant branch (super--AGB) star. The estimated mass of the progenitor is close to the upper limit of the mass range typical of this class of stars, M$_{mas}$ (see \citealt{Pumo2009ECSN} and references therein). This seems to corroborate the results of \cite{PumoModel2017}, showing that some faint SNe IIP may be also explained in terms of ECSNe involving massive super--AGB stars. 
To investigate this scenario in more detail, we compare the photometric and spectroscopic properties of SN 2020cxd with other ECSN candidates in Appendix \ref{AppendixECSN}.
We note, however, that we lack conclusive evidence to confidently discriminate between an ECSN scenario and a standard faint SN IIP event with a RSG progenitor.


\subsection{SN 2021aai results and progenitor scenarios}
We also perform hydrodynamic modelling of SN 2021aai in the high reddening scenario, assuming it is the most reliable of the two (Figure \ref{HydroAAI}). We obtained R = 4 $\times$ 10$^{13}$ cm ($\sim$ 575 R$_{\odot}$), M$_{ej}$ = 15.5 M$_{\odot}$, and E = 0.4 foe. Given the higher energy and ejected mass compared to SN 2020cxd, we favour the scenario where a RSG explodes through an iron core collapse, excluding the ECSN origin for SN 2021aai.
In Figure \ref{E/M2} it is possible to appreciate that SN 2021aai belongs to the category labelled as ``intermediate-luminosity'' SNe \citep{PumoModel2017,Tomasella2018}, which bridge the classes of LL SNe IIP and standard SN IIP events, therefore creating a continuous distribution in the properties of SN IIP.
As we remark in Sect. \ref{DiscoveryPhotometry}, this transient is characterized by an extended plateau phase, lasting $\sim$140 days. This feature is not well reproduced by our hydrodynamical model, which predicts a shorter plateau compared to observations (Figure \ref{HydroAAI}, top panel). This difference between the model and the observations could be probably explained in terms of  a peculiar distribution of the \textsuperscript{56}Ni within the ejected material. In fact, keeping constant the basic parameters of the models (i.e. M$_{ej}$, R, E  and the total amount of \textsuperscript{56}Ni initially present in the ejected envelope), different degrees of \textsuperscript{56}Ni mixing primarily lead to different plateau durations (see e.g. Figure 11 in \citealt{Pumo_Mixing_Ni}). In particular a lower degree of \textsuperscript{56}Ni mixing (i.e. models where the Ni is more confined to the central region of the ejecta) is linked to a longer plateau, as observed for SN 2020cxd.
We also perform some preliminary hydrodynamical modelling of SN 2021aai in the low reddening scenario. Firstly, we notice that the plateau temperature of 4300 K was too low to be fitted by our models, making the high reddening a more reliable scenario. Fitting only the bolometric light curve and the expansion velocities, we obtain values of R and E reduced by a factor of $\sim$1.5--2 and a ratio E/M$_{ej}$ almost unchanged compared to the high reddening scenario.

\section{Summary and conclusions} \label{Concl}
We present optical photometry and spectroscopy for two LL SNe IIP: 2020cxd and 2021aai.
SN 2020cxd appears to be sub--luminous even compared to other transients in its class, with an absolute magnitude of M$_{r}$ = --14 mag at the start of the plateau, making it one of the faintest LL SNe IIP observed to date. On the other hand, SN 2021aai is a transitional object between LL SNe IIP and more standard SN IIP events, once corrected for the large extinction affecting the target ($A_{V}$ = 1.9 mag). Both transients display spectra that perfectly match those of other LL SNe IIP \citep{PastoLLSNe2004,SpiroPasto2014}, characterised by H lines in the early phases and followed by the rise of metal lines (mainly Fe II, Sc II, Ba II, [Ca II] and Ca NIR triplet) during the plateau phase. The expansion velocities obtained by measuring the position of the minimum of the P Cygni line profile, well visible for most lines, yields velocities of few 10$^{3}$ km s$^{-1}$, below those of standard SNe IIP, but in line with what was observed for LL SNe IIP.
The temperature trend obtained through spectral energy distribution fitting consists in a very rapid decline during the early phases, reaching a temperature of $\sim$5500 K at $\sim$30 days after explosion and throughout all the plateau phase, as expected for H recombination.
After fading from the plateau, both objects settle on the linear decline powered by the \textsuperscript{56}Ni decay chain. By comparing their late time luminosity with that of SN 1987A at the same phase, we estimate the \textsuperscript{56}Ni synthesised to be 1.8 $\pm$ 0.5 $\times$ 10$^{-3}$ M$_{\odot}$ for SN 2020cxd and 1.4 $\pm$ 0.5 $\times$ 10$^{-2}$ M$_{\odot}$ for SN 2021aai (considering the high reddening scenario).

We also perform hydrodynamical modelling of our targets using the procedure described in \cite{PumoModel2017}, which uses the general--relativistic, radiation--hydrodynamics, Lagrangian code presented in \cite{PumoEZampieriModel}. The physical parameters of the progenitor star of SN 2021aai at the moment of explosion are R = 4 $\times$ 10$^{13}$ cm ($\sim$ 575 R$_{\odot}$), M$_{ej}$ = 15.5 M$_{\odot}$ and $E$ = 0.4 foe. These values are consistent with the explosion of a RSG star after the collapse of its iron core \citep{WheelerFeCC}. The transitional properties of SN 2021aai, linking LL SNe IIP and standard SN IIP events, are evident when considering its E/M$_{ej}$ ratio (Figure \ref{E/M2}).
The interpretation of the parameters obtained for SN 2020cxd is more nuanced. The best fit yields R = 4 $\times$ 10$^{13}$ cm ($\sim$ 575 R$_{\odot}$), M$_{ej}$ = 7.5 M$_{\odot}$ and $E$ = 0.097 foe, values which can be compatible with the iron core collapse explosion of a low mass (8.9--10.4 M$_{\odot}$) RSG, but they are also consistent with an explosion triggered by electron captures involving a massive super--AGB (i.e. with an initial mass close to the upper limit of the mass range typical of this class of stars, M$_{mas}$; see \citealt{Pumo2009ECSN}, and references therein).
In conclusion, we analyse two objects spanning the brightest and faintest edges of the LL SNe IIP class, with SN 2021aai bridging the low luminosity class with more traditional SNe IIP, and SN 2020cxd being so faint that it can be reasonably considered a possible ECSN candidate.


\section*{Acknowledgements}

We thank the staffs of the various observatories where data were obtained for their assistance.
Based on observations made with the Nordic Optical Telescope, owned in collaboration by the University of Turku and Aarhus University, and operated jointly by Aarhus University, the University of Turku and the University of Oslo, representing Denmark, Finland and Norway, the University of Iceland and Stockholm University at the Observatorio del Roque de los Muchachos, La Palma, Spain, of the Instituto de Astrofisica de Canarias.
Observations from the Nordic Optical Telescope were obtained through the NUTS2 collaboration which are supported in part by the Instrument Centre for Danish Astrophysics (IDA). The data presented here were obtained in part with ALFOSC, which is provided by the Instituto de Astrofisica de Andalucia (IAA).
This work makes use of data from the Las Cumbres Observatory network. The LCO team is supported by NSF grants AST--1911225 and AST--1911151, and NASA SWIFT grant 80NSSC19K1639. 
Data were also obtained at the Liverpool Telescope, which is operated on the island of La Palma by Liverpool John Moores University in the Spanish Observatorio del Roque de los Muchachos with financial support from the UK Science and Technology Facilities Council. Part of the observations were collected at Copernico and Schmidt telescopes (Asiago, Italy) of the INAF – Osservatorio Astronomico di Padova. This work was based in part on observations made with the Italian Telescopio Nazionale Galileo (TNG) operated on the island of La Palma by the Fundaci\'{o}n Galileo Galilei of the INAF (Istituto Nazionale di Astrofisica) at the Spanish Observatorio del Roque de los Muchachos of the Instituto de Astrofisica de Canarias.
Based on observations made with the Gran Telescopio Canarias (GTC), installed in the Spanish Observatorio del Roque de los Muchachos of the Instituto de Astrofísica de Canarias, in the island of La Palma
This work has made use of data from the Asteroid Terrestrial-impact Last Alert System (ATLAS) project. 
M.L.P. acknowledges support from the plan ``programma ricerca di ateneo UNICT 2020-22 linea 2'' of the University of Catania.
N.E.R. acknowledges partial support from MIUR, PRIN 2017 (grant 20179ZF5KS), from the Spanish MICINN grant PID2019-108709GB-I00 and FEDER funds, and from the program Unidad de Excelencia María de Maeztu CEX2020-001058-M.
L.G. acknowledges financial support from the Spanish Ministerio de 
Ciencia e Innovaci\'on (MCIN), the Agencia Estatal de Investigaci\'on 
(AEI) 10.13039/501100011033, and the European Social Fund (ESF) 
"Investing in your future" under the 2019 Ram\'on y Cajal program 
RYC2019-027683-I and the PID2020-115253GA-I00 HOSTFLOWS project, from 
Centro Superior de Investigaciones Cient\'ificas (CSIC) under the PIE 
project 20215AT016, and the program Unidad de Excelencia Mar\'ia de 
Maeztu CEX2020-001058-M.
T.M.B. acknowledges financial support from the Spanish 
Ministerio de Ciencia e Innovaci\'on (MCIN), the Agencia Estatal de 
Investigaci\'on (AEI) 10.13039/501100011033 under the 
PID2020-115253GA-I00 HOSTFLOWS project, and from Centro Superior de 
Investigaciones Cient\'ificas (CSIC) under the PIE project 20215AT016, 
and the program Unidad de Excelencia Mar\'ia de Maeztu CEX2020-001058-M.
Y.-Z. Cai is funded by China Postdoctoral Science Foundation (grant no. 2021M691821)




\bibliographystyle{mnras}
\bibliography{example} 



\clearpage
\newpage

\appendix
\section{Comparison between ECSN candidates} \label{AppendixECSN}
\FloatBarrier

Given the possibility of SN 2020cxd originating from an ECSN scenario, as highlighted in Sect. \ref{cxdECSN}, in this appendix we present a comparison between SN 2020cxd and other ECSN candidates.
The first object we selected for this purpose is the peculiar type II SN 2018zd \citep{Hiramatsu2018zd}.
\cite{Hiramatsu2018zd} found several indicators favouring the ECSN event for this transient, in particular the chemical composition of the progenitor and the results of the nucleosynthesis, the light curve morphology and the presence of CSM.
We also chose to include in this small sample SN 2008S \citep{Botticella2008S}, taken as a prototype of ILRTs. This class was associated to ECSNe due to their faintness (e.g. \citealt{Hump2011}), their progenitors \citep{ProgenitorDetection,Thompson2008} and the presence of circumstellar material, clearly evident in all their spectra, which corroborates their origin from a Super-AGB progenitor.

\begin{figure}
\begin{center}
\includegraphics[width=1\columnwidth]{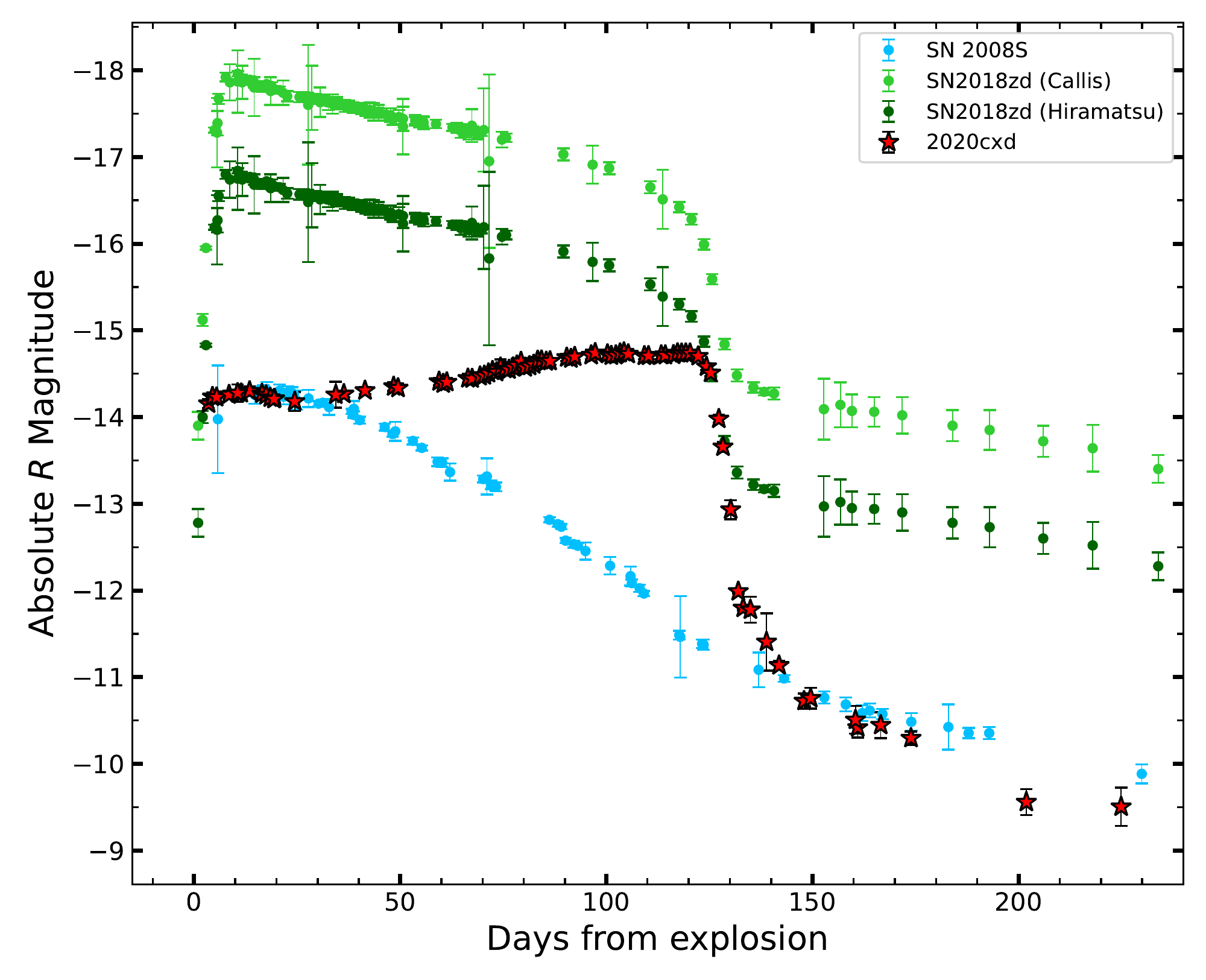}

\caption{Light curve comparison between SN 2020cxd and two other ECSN candidates: the ILRT SN 2008S and the peculiar SN 2018zd. See text for details.} \label{AppendixLC}
\end{center}

\end{figure}

In Figure \ref{AppendixLC}, we show the $R$ band (correction between $R$ and $r$ bands were applied as discussed in Sect. \ref{r-R}) light curves of the three transients mentioned. For SN 2018zd we adopt both distances reported in \cite{Hiramatsu2018zd} and \cite{Callis2018zd}.
The increase in brightness during the plateau of SN 2020cxd is striking, since it is the only object displaying this behaviour. SN 2018zd shows perhaps a more canonical plateau, slightly declining in brightness over the course of $\sim$ 120 days.
The late time decline of SN 2008S is almost coincident with that of SN 2020cxd.

\begin{figure}
\begin{center}

\includegraphics[width=1\columnwidth]{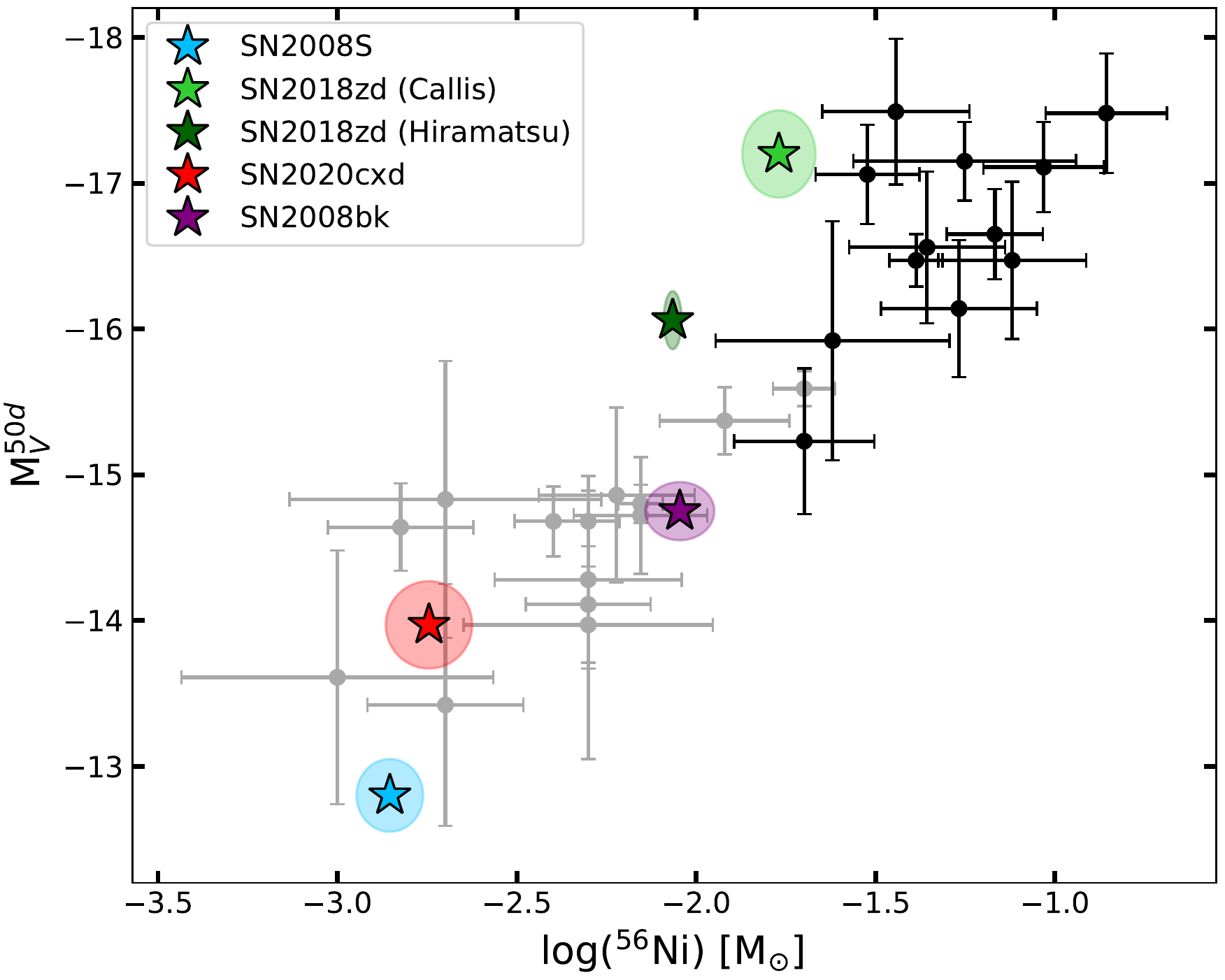}
\caption{$V$ band absolute magnitude at 50 days versus \textsuperscript{56}Ni ejected mass. LL SNe are shown in grey, standard SN IIP are represented in black. SN 2020cxd, SN 2018zd and SN 2008S are highlighted with coloured stars, with their errors reported as elliptical regions.} \label{AppendixNi}
\end{center}

\end{figure}

To better visualize the relationship between these objects and the data shown in Sect. \ref{NiEst}, in Figure \ref{AppendixNi} we plot SN 2008S and SN 2018zd on the M$_{V}$ - \textsuperscript{56}Ni diagram already shown in Figure \ref{ConfNiMag}. We note that a tight relationship between these two quantities was found for SNe IIP, but SN 2008S fits remarkably well in the lower end of the brightness distribution despite being a member of a different class of transients. SN 2018zd appears to belong to the standard IIP events when correcting for the distance reported by \cite{Callis2018zd}, while it lies towards the region of transitional objects (like SN 2021aai) when adopting the distance prescribed by \cite{Hiramatsu2018zd}.

\begin{figure*}
\begin{center}

\includegraphics[width=1.9\columnwidth]{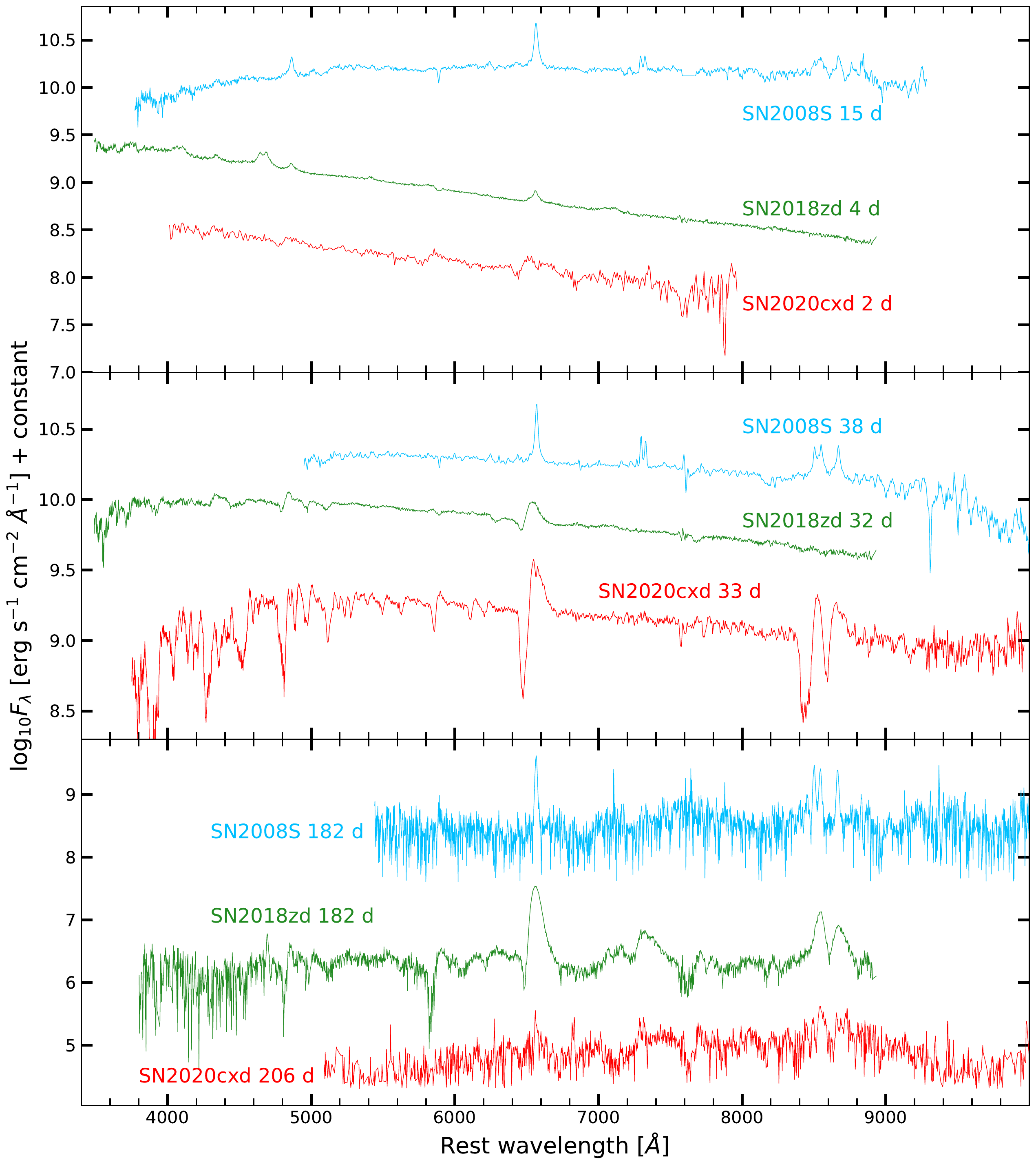}

\caption{Comparison between the spectra of SN 2008S, SN 2018zd and SN 2020cxd. In the top panel are shown the early spectra, in the middle panel are presented the spectra during the plateau phase, and in the bottom panel are shown the late spectra. All spectra were corrected for redshift and reddening.} \label{AppendixSpec}
\end{center}

\end{figure*}

In Figure \ref{AppendixSpec} we present a comparison between the spectra of this small sample of ECSN candidates at early, middle and late phases (top, middle and bottom panel respectively).
The earliest phase available for a spectrum of SN 2008S is at 15 days after the explosion, while the spectra of SNe 2018zd and 2020cxd were taken within 4 days from the explosion. However, we note that the spectra of ILRTs evolve very slowly, due to being dominated by CSM emission: for this reason, the main features characterizing the spectra do not change on short time scales.
In the early spectrum of SN 2018zd it is possible to notice some narrow H lines without P-Cygni profiles, somehow reminiscent of the ILRT spectrum (although with a much bluer colour), which can be traced back to the presence of CSM, as \cite{Hiramatsu2018zd} infer from their analysis of the ultraviolet colour evolution. At the same phases, SN 2020cxd already shows P-Cygni profiles and broad H lines, in line with the expectations for a LL SN IIP.
At $\sim$ 30-40 days, SN 2008S shows almost no sign of evolution, with the narrow H and Ca lines completely dominating the spectrum. SN 2020cxd, on the other hand, develops an abundance of metal lines (the line-blanketing effect is already evident) and deep P-Cygni profiles. In this phase SN 2018zd transitions towards a more standard SN IIP, although the metal lines are still much weaker compared to SN 2020cxd, the line-blanketing effect is not marked, and some signature features such as the Ca NIR triplet are still missing.
Finally, at late times the spectrum of SN 2008S has kept basically the same narrow lines it has shown throughout its evolution, even without a continuum underneath them. SN 2018zd displays an array of prominent emission lines, allowing the detailed analysis performed by \cite{Hiramatsu2018zd} which stated that this object is compatible with an ECSN event on the basis of the nucleosynthesis and chemical composition expectations\footnote{Despite the presence of prominent emission lines, we note that \cite{Hiramatsu2018zd} and \cite{Callis2018zd} disagree on the presence of abundance patterns that support the ECSN hypothesis.}.
Sadly, it was impossible to perform a similar analysis on SN 2020cxd, due to the very poor signal to noise obtained in our latest spectra.

In conclusion, applying the criteria presented by \cite{Hiramatsu2018zd} to identify an ECSN event, we notice the following pros and cons:
\begin{itemize}
    \item The low energy characterizing SN 2020cxd and its light curve shape are consistent with an ECSN origin. As shown in Sect. \ref{cxdECSN}, hydrodynamical modelling points towards a progenitor between 8.9 and 10.4 M$_{\odot}$, compatible with the expectations for a super--AGB star. 
    \item We did not have any direct detection of the progenitor, nor we could investigate the nucleosynthesis and chemical composition of the progenitor through nebular spectra. 
    \item The lack of CSM that can be inferred from the spectra seems to point towards a low--mass Red Giant Branch (RGB) progenitor, rather than a super-AGB, therefore favouring the iron core--collapse scenario for SN 2020cxd (although a Super-AGB star could explode without being surrounded by optically thick CSM in some cases, see e.g \citealt{Pumo2009ECSN}).
\end{itemize}

We know that low--mass RGB progenitors were accurately identified in the past, e.g. for SN 2008bk (\citealt{Dyk2008bk,Maund2014,ONeill2021}) and SN 2018aoq (\citealt{Progen2018aoq}). A similar scenario could comfortably explain the SN 2020cxd event.


\section{Supplementary tables and pictures} \label{data_tables}

\begin{figure}
\makebox[\textwidth][c]{\includegraphics[width=1.3\columnwidth]{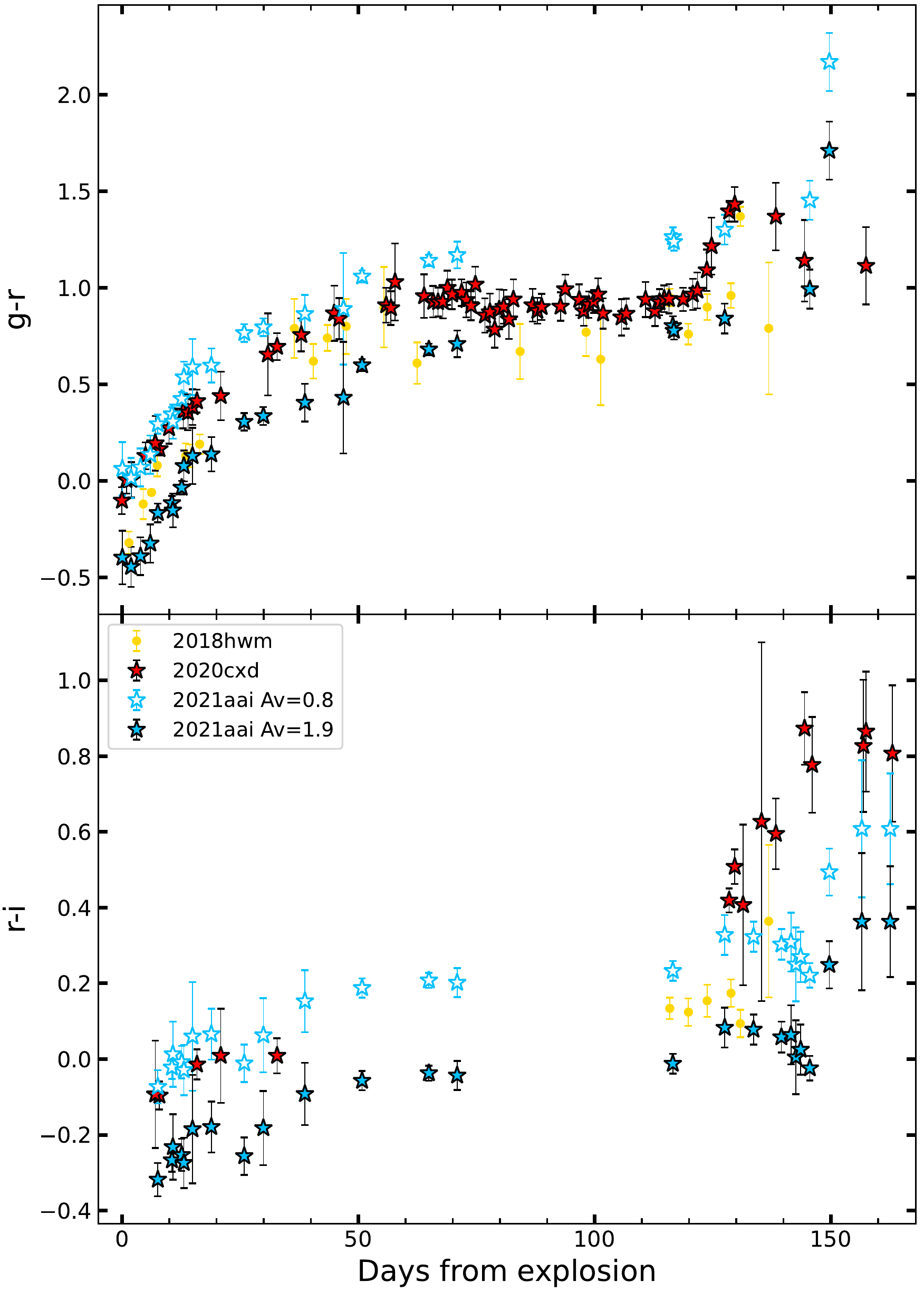}}
\caption{$g-r$ and $r-i$ colour evolution for SNe 2018hwm, 2020cxd and 2021aai.} \label{Colour_appenidx}

\end{figure}


\begin{table*}
	\centering
	\caption{List of instruments and facilities used in our follow--up campaigns, detailing also the filters used to take photometric data. See Table \ref{LogSpectra} for details about the spectra.} 
	\label{Instrum}
	\begin{tabular}{lllll} 
		\hline
Code & Telescope, [m] & Instrument & Filters & Site \\
		\hline
EKAR &  Schmidt, 0.91 & Moravian & V,g,r,i & Osservatorio Astronomico di Asiago, Cima Ekar \\
fl03--fl15 &  LCO\textsuperscript{$\dagger$} (LSC site), 1.00 &  Sinistro & U,B,V,g,r,i,z & Cerro Tololo Inter--American Observatory\\
fl05--fl07 &  LCO (ELP site), 1.00 &  Sinistro & U,B,V,g,r,i,z &  McDonald Observatory\\
fl06--fl14 &  LCO (CPT site), 1.00 &  Sinistro & U,B,V,g,r,i,z & South African Astronomical Observatory\\
fl12 &  LCO (COJ site), 1.00 &  Sinistro & U,B,V,g,r,i,z &  Siding Spring Observatory\\
ZTF  &  Oschin Telescope, 1.22  & ZTF & g,r &  Palomar Observatory, United States \\
AFOSC &  Copernico Telescope, 1.82 & AFOSC & B,V,g,r,i,z & Osservatorio Astronomico di Asiago, Cima Ekar\\
IO:O &  Liverpool Telescope, 2.00 & IO:O & B,V,g,r,i,z & Observatorio Roque de Los Muchachos, La Palma\\
FLOYDS &  LCO (FTN/FTS site), 2.00  & FLOYDS & -- & Haleakala (FTN) and Australia (FTS)\\
ALFOSC &  Nordic Optical Telescope, 2.56  & ALFOSC & B,V,g,r,i,z & Observatorio Roque de Los Muchachos, La Palma\\
NOTCam & Nordic Optical Telescope, 2.56   & NOTCam & J,H,K & Observatorio Roque de Los Muchachos, La Palma\\
LRS &  Telescopio Nazionale Galileo, 3.58  & DOLORES & B,V,u,g,r,i,z & Observatorio Roque de Los Muchachos, La Palma\\
OSIRIS &  Gran Telescopio CANARIAS, 10.40  & OSIRIS & -- & Observatorio Roque de Los Muchachos, La Palma\\
		\hline
		\textsuperscript{$\dagger$} Las Cumbres Observatory
	\end{tabular}
\end{table*}




\begin{table*}
\centering
\caption{Photometric data in the Sloan filters collected for SN 2020cxd (AB mag).}
\label{tab1}
\begin{tabular}{ c c c c c c c } 
 \hline
 Date & MJD & g  & r  & i & z & Instrument  \\ 
 \hline
 \hline
  & & & & & & \\

2020/02/26 & 58905.58 &  17.80  0.1   & 17.57   0.10 &  17.64   0.10 &  --   & LCO	   \\
2020/02/27 & 58906.46 &  17.77  0.02  & 17.57   0.03 &  17.65   0.03 &  --   & LCO 	   \\
2020/03/06 & 58914.41 &  18.09  0.05  & 17.64  0.03  &  17.63   0.03 &  --   & LCO	   \\
2020/03/11 & 58919.44 &  18.14  0.06  & 17.67  0.11  &  17.64   0.06 &  --   & LCO	   \\
2020/03/21 & 58929.40 &  18.28  0.15  & 17.59   0.15 &  --  & 17.42   0.15  & LCO	   \\
2020/03/23 & 58931.36 &  18.31  0.06  & 17.58   0.04 &  17.55   0.03 &  --   & LCO 	   \\
2020/05/09 & 58978.57 &  --  &  -- &  16.98   0.01 &  --   & Pan-STARRS		   \\
2020/05/15 & 58983.56 &  --  &  -- &  16.95   0.01 &  --   & Pan-STARRS		   \\
2020/06/10 & 59010.57 &  --  &  -- &  16.93   0.02 &  --   & Pan-STARRS		   \\
2020/06/14 & 59014.51 &  --  &  -- &  16.92   0.01 &  --   & Pan-STARRS		   \\
2020/06/27 & 59027.03 &  21.29  0.05  & 19.86   0.02 &  19.42  0.02 & 19.07   0.03  & IO:O   \\
2020/06/27 & 59027.40 &  --  &  -- &  19.46   0.11 &  --  & Pan-STARRS		   \\
2020/06/28 & 59028.22 &  21.52  0.09  & 20.05   0.03 &  19.52   0.04 & 19.21    0.05  & IO:O   \\
2020/06/29 & 59029.97 &  --   & 20.07   0.15 &  19.64   0.15 & 19.39    0.15  & OSIRIS	   \\
2020/07/06 & 59036.93 &  22.12  0.17  & 20.71   0.05 &  20.09   0.08 & 19.67    0.09  & IO:O   \\
2020/07/12 & 59042.99 &  22.30  0.19  & 21.12   0.09 &  20.23   0.04 & 19.82    0.10  & IO:O   \\
2020/07/25 & 59055.99 &  22.58  0.17  & 21.43   0.11 &  20.54   0.11 & 19.98    0.09  & IO:O   \\
2020/07/30 & 59060.46 &  --  &  -- &  20.57   0.29 &  --  & Pan-STARRS		   \\
2020/08/01 & 59062.50 &  --  &  -- &  20.61   0.14 &  --  & Pan-STARRS		   \\
2020/08/07 & 59068.93 &  --   & 21.55   0.08 &  20.74  0.06 & 20.07    0.07  & IO:O 	   \\
2020/08/18 & 59079.01 &  --   &  --   &  21.08   0.05 & 20.39    0.05  & ALFOSC	   \\
2020/09/04 & 59096.91 &  --   & 22.29   0.15 &  21.71   0.17 & 20.75    0.16  & IO:O	   \\
2020/09/27 & 59119.90 &  --   & 22.34   0.22 &  21.74   0.09 & 21.24    0.14  & IO:O	   \\
2020/09/30 & 59122.84 &  --   &  >22.02  &  21.73   0.09 & 21.32   0.14  & IO:O 		   \\
 & & & & & & \\
 \hline

\end{tabular}
\end{table*}

\begin{table*}
\centering
\caption{Photometric data collected with Johnson filters for SN 2020cxd (Vega mag).}
\label{tab2}
\begin{tabular}{ c c c c c c  } 
 \hline
 Date & MJD & U & B & V  & Instrument   \\ 
 \hline
 \hline
 & & & & &   \\

2020/02/26  & 58905.58 &   --  &  17.98   0.10 &  17.73   0.10  & LCO     \\
2020/02/27  & 58906.44 &  17.41   0.05 &  18.06   0.03  & 17.71   0.03  & LCO     \\
2020/03/06  & 58914.39 &  18.57   0.10 &  18.43   0.07  & 17.81   0.05  & LCO     \\
2020/03/11  & 58919.43 &  18.86   0.19 &  18.67   0.07  & 17.90   0.06  & LCO     \\
2020/03/21  & 58929.30 &   -- &  18.97   0.20 &  17.85   0.15  & LCO     \\
2020/03/23  & 58931.35 &  19.53   0.15 &  18.94  0.09 &  17.84   0.04  & LCO     \\
2020/06/28  & 59029.97 &   --  &   --  &  20.93   0.20  & OSIRIS  \\

 & & & & &  \\
 \hline

\end{tabular}
\end{table*}

\begin{table*}
\centering
\caption{Photometric data collected through the ATLAS survey for SN 2020cxd (AB mag).}
\label{tab3}
\begin{tabular}{ c c c c c  } 
 \hline
 Date & MJD & cyan  & orange  & Instrument   \\ 
 \hline
 \hline
 & & & &    \\

2020/03/02 &   58910.64   &	--   &  17.56  0.17	&  ATLAS  \\
2020/03/26 &   58934.63   &	--   &  17.42  0.23	&  ATLAS  \\
2020/03/30 &   58938.62   &	--   &  17.57  0.10	&  ATLAS  \\
2020/04/03 &   58942.63   &	--   &  17.54  0.20	&  ATLAS  \\
2020/04/07 &   58946.61   &	--   &  17.46  0.38	&  ATLAS  \\
2020/04/11 &   58950.63   &	--   &  17.50  0.03	&  ATLAS  \\
2020/04/17 &   58956.59   &	--   &  17.38  0.08	&  ATLAS  \\
2020/04/21 &   58960.58   &	17.84   0.11  &   --   &  ATLAS  \\
2020/04/23 &   58962.60   &	--   &  17.33   0.05	&  ATLAS  \\
2020/04/25 &   58964.53   &	17.82   0.13  &   --   &  ATLAS  \\
2020/04/27 &   58966.52   &	--   &  17.26   0.07	&  ATLAS  \\
2020/04/29 &   58968.49   &	17.70   0.02  &   --   &  ATLAS  \\
2020/05/01 &   58970.52   &	--   &  17.24  0.09	&  ATLAS  \\
2020/05/03 &   58972.52   &	--   &  17.22  0.03	&  ATLAS  \\
2020/05/05 &   58974.52   &	--   &  17.21  0.09	&  ATLAS  \\
2020/05/07 &   58976.48   &	--   &  17.18  0.06	&  ATLAS  \\
2020/05/09 &   58978.50   &	--   &  17.14  0.04	&  ATLAS  \\
2020/05/13 &   58982.54   &	--   &  17.01  0.09	&  ATLAS  \\
2020/05/15 &   58984.56   &	--   &  17.18  0.04	&  ATLAS  \\
2020/05/17 &   58986.55   &	--   &  17.14  0.05	&  ATLAS  \\
2020/05/19 &   58988.50   &	17.56   0.10  &   --   &  ATLAS  \\
2020/05/21 &   58990.45   &	--   &  17.05  0.01	&  ATLAS  \\
2020/05/23 &   58992.44   &	17.55   0.09  &   --   &  ATLAS  \\
2020/05/25 &   58994.47   &	--   &  17.06  0.11	&  ATLAS  \\
2020/05/29 &   58998.46   &	--   &  17.07  0.03	&  ATLAS  \\
2020/05/31 &   59000.52   &     17.53   0.04  &   --   &  ATLAS  \\
2020/06/02 &   59002.51   &	--   &  17.08  0.12	&  ATLAS  \\
2020/06/04 &   59004.59   &	--   &  17.05  0.30	&  ATLAS  \\
2020/06/06 &   59006.41   &	--   &  17.03  0.06	&  ATLAS  \\
2020/06/08 &   59008.51   &	--   &  17.06  0.36	&  ATLAS  \\
2020/06/10 &   59010.42   &	--   &  17.13  0.04	&  ATLAS  \\
2020/06/11 &   59011.47   &	--   &  17.14  0.06	&  ATLAS  \\
2020/06/14 &   59014.41   &	--   &  17.17  0.13	&  ATLAS  \\
2020/06/15 &   59015.41   &	17.57   0.10  &    --    &  ATLAS  \\
2020/06/18 &   59018.43   &	--   &  17.18  0.10	&  ATLAS  \\
2020/06/20 &   59020.45   &	17.91  0.13  &   --   &  ATLAS  \\
2020/06/28 &   59028.42   &	>20.41    & >19.93  &  ATLAS  \\
2020/07/18 &   59048.41   &	>20.38  &   --   &  ATLAS  \\
2020/07/20 &   59050.37   &	--   & >20.25  &  ATLAS  \\

 & & & &    \\
 \hline
\end{tabular}
\end{table*}

\begin{table*}
\centering
\caption{Photometric data in the Sloan filters collected for SN 2021aai (AB mag).}
\label{tab4}
\begin{tabular}{ c c c c c c c c  } 
 \hline
 Date & MJD & u & g & r & i  & z & Instrument  \\ 
 \hline
 \hline
 & & & & & & &  \\

2021/01/08 & 59222.27 &    -- &   --      &  >20.54     &   --   &   --  & ZTF 	    \\
2021/01/10 & 59224.42 &    -- & 18.85   0.08    &  --     &   --   &   --  & ZTF  	    \\
2021/01/12 & 59226.40 &	   -- &   18.76    0.11	&  18.40     0.09   &   --	  &  --    & ZTF	    \\	
2021/01/14 & 59228.24 &	   -- &   18.62    0.08	&  18.32     0.07   &   --	  &  --    & ZTF	    \\
2021/01/16 & 59230.24 &	   -- &   18.66    0.08	&  18.30     0.06   &   --	  &  --    & ZTF	    \\
2021/01/18 & 59232.32 &	   -- &   18.72    0.09	&  18.29     0.05   &   --	  &  --    & ZTF	    \\
2021/01/19 & 59233.91 &  19.76  0.04 & 18.90    0.04   &  18.32       0.03 &  18.26  0.03  & 18.15  0.05   & LRS          \\
2021/01/22 & 59236.90 &  20.20  0.25 & 18.97    0.030   &  18.34       0.02 &  18.23  0.02  & 18.11  0.03   & ALFOSC  \\
2021/01/23 & 59237.09 &    -- & 18.96    0.07   &  18.37       0.06 &  18.22  0.06  &   --    & LCO	    \\
2021/01/24 & 59238.96 &  20.68  0.15 & 19.01     0.01   &  18.29       0.03 &  18.17  0.03  & 18.01  0.04   & ALFOSC  \\
2021/01/25 & 59239.41 &    -- & 19.14     0.07   &  18.32       0.04 &  18.21  0.05  &   --    & LCO 	    \\
2021/01/27 & 59241.26 &    -- & 19.21     0.12   &  18.33       0.09 &  18.14  0.11  &   --    & LCO	    \\
2021/01/31 & 59245.23 &    -- & 19.22     0.07   &  18.34       0.05 &  18.14  0.05  &   --    & LCO	    \\
2021/02/03 & 59248.17 &    -- & 19.31     0.04   &  --  &  18.15  0.04  &   --    & LCO	    \\
2021/02/03 & 59248.39 &    -- & 19.36    0.11   &  18.31     0.07   &   --   &   --  & ZTF 	    \\
2021/02/05 & 59250.22 &	   -- &  19.31    0.18	&  18.28     0.07   &   --	  &  --    & ZTF	    \\
2021/02/07 & 59252.19 &	   -- &   19.36    0.12	&  18.26     0.05   &   --	  &  --    & ZTF	    \\
2021/02/07 & 59252.20 &    -- & 19.31     0.04   &  18.25       0.03 &  18.13  0.04  &   --    & LCO	    \\
2021/02/09 & 59254.18 &	   -- &   19.32    0.14	&  18.23     0.05   &   --	  &  --    & ZTF	    \\	
2021/02/11 & 59256.18 &	   -- &   19.28    0.11	&  18.26     0.05   &   --	  &  --    & ZTF	    \\
2021/02/11 & 59256.25 &    -- & 19.35     0.03   &  18.26       0.03 &  18.07  0.09  &   --    & LCO	    \\
2021/02/15 & 59260.39 &	   -- &  19.31    0.21	&  18.22     0.08   &   --	  &  --    & ZTF	    \\
2021/02/18 & 59263.21 &	   -- &   --	&  18.19     0.10   &   --	  &  --    & ZTF	    \\	
2021/02/20 & 59265.04 &    -- & 19.28     0.08   &  18.13       0.06 &  17.85  0.06  & 17.93  0.18   & ALFOSC  \\
2021/02/20 & 59265.31 &	   -- &   19.36    0.13	&  18.15     0.07   &   --	  &  --    & ZTF	    \\	
2021/02/22 & 59267.30 &	   -- &   19.36    0.19	&  18.10     0.08   &   --	  &  --    & ZTF	    \\
2021/02/26 & 59271.23 &    -- &  --    &  18.12       0.08 &    --  &   --    & LCO	    \\
2021/02/26 & 59271.27 &    -- &   --      &  18.14     0.08   &   --   &   --  & ZTF 	    \\
2021/02/28 & 59273.21 &    -- & 19.32    0.27    &  18.13     0.10   &   --   &   --  & ZTF 	    \\
2021/03/02 & 59275.33 &	   -- &   --	&  18.07    0.06   &   --	  &  --    & ZTF	    \\	
2021/03/04 & 59277.13 &    -- & 19.47     0.03   &  18.12       0.02 &  17.80  0.02  &   --    & LCO 	    \\
2021/03/05 & 59278.18 &	   -- &   19.33    0.18	&  18.06     0.07   &   --	  &  --    & ZTF	    \\
2021/03/08 & 59281.21 &	   -- &  19.42    0.13	&  18.06     0.06   &   --	  &  --    & ZTF	    \\
2021/03/14 & 59287.06 &    -- & --   &  18.20  0.24 &  17.85  0.11  &   17.60 0.05    & AFOSC 	    \\
2021/03/18 & 59291.20 &    -- &   --      &  18.10     0.07   &   --   &   --  & ZTF 	    \\
2021/03/18 & 59291.27 &    -- & 19.50     0.03   &  18.07       0.02 &  17.74 0.01  &   --    & LCO	    \\
2021/03/20 & 59293.21 &    -- &   19.53    0.11     &  18.09     0.08   &   --   &   --  & ZTF 	    \\
2021/03/23 & 59296.22 &    -- &   --      &  17.98     0.13   &   --   &   --  & ZTF 	    \\
2021/03/24 & 59297.25 &    -- & 19.53   0.06   &  18.07       0.03 &  17.74  0.03  &   --    & LCO 	    \\
2021/03/25 & 59298.29 &    -- &   --      &  18.02     0.22   &   --   &   --  & ZTF 	    \\
2021/03/29 & 59302.17 &	   -- & 19.58    0.22	&  --    &   --	  &  --    & ZTF	    \\
2021/03/31 & 59304.16 &    -- &   --      &  17.98   0.06   &   --   &   --  & ZTF 	    \\

 & & & & & & &   \\
 \hline
\end{tabular}
\end{table*}

\setcounter{table}{4}
\renewcommand{\thetable}{B\arabic{table}}

\begin{table*}
\centering
\caption{\textbf{(Continued)} Photometric data in the Sloan filters collected for SN 2021aai (AB mag).}
\label{tab5}
\begin{tabular}{ c c c c c c c c  } 
 \hline
 Date & MJD & u & g & r & i  & z & Instrument  \\ 
 \hline
 \hline
 & & & & & & & \\

2021/04/02 & 59306.31 &	   -- & 19.34    0.14	&  17.97     0.09   &   --	  &  --    & ZTF	    \\
2021/04/04 & 59308.22 &    -- &   19.42    0.15     &  17.97     0.06   &   --   &   --  & ZTF	    \\
2021/04/06 & 59310.23 &	   -- & 19.33    0.12	&  --    &   --	  &  --    & ZTF	    \\
2021/04/08 & 59312.28 &	   -- &   19.36    0.14	&  17.94     0.06 &   --	  &  --    & ZTF	    \\
2021/04/10 & 59314.28 &	   -- &   -- 	&  17.89      0.05 &   --	  &  --    & ZTF	    \\
2021/04/12 & 59316.28 &	   -- &   19.32    0.11	&  17.83       0.04 &   --	  &  --    & ZTF	    \\	
2021/04/15 & 59319.27 &	   -- & 19.34    0.13	&  --    &   --	  &  --    & ZTF	    \\
2021/04/18 & 59322.27 &	   -- &   19.31    0.19	&  17.79       0.06 &   --	  &  --    & ZTF	    \\	
2021/04/20 & 59324.22 &	   -- &   19.25    0.15	&  17.74       0.06 &   --	  &  --    & ZTF	    \\
2021/04/24 & 59328.18 &	   -- &   --	&  17.70       0.06 &  --	  &  --    & ZTF	    \\	
2021/05/08 & 59342.89 &    -- & 19.27     0.05   &  17.72       0.02 &  17.36  0.01  & 17.15 0.02   & ALFOSC  \\
2021/05/09 & 59343.12 &    -- & 19.26     0.04   &  17.74       0.02 &    --  &   --    & LCO 	    \\
2021/05/19 & 59353.89 &    -- & 19.33     0.07   &  17.74       0.04 &  17.28  0.03  &   --    & Moravian	    \\
2021/05/25 & 59359.97 &    -- &   --   &  17.75       0.03 &  17.30  0.03  &   --    & Moravian	    \\
2021/05/31 & 59365.87 &    -- &   --   &  17.81       0.03 &  17.38  0.03  &   --    & Moravian	    \\
2021/06/02 & 59367.89 &    -- &   --   &  17.96       0.04 &  17.52  0.07  &   --    & Moravian	    \\
2021/06/03 & 59368.94 &    -- &   --   &  18.07       0.06 &  17.69  0.08  &   --    & Moravian     \\
2021/06/04 & 59369.95 &    -- &   --   &  18.14       0.06 &  17.74  0.03  &   --    & Moravian	    \\
2021/06/06 & 59371.89 &    -- & 20.10     0.10   &  18.36       0.03 &  18.01  0.02  & 17.75  0.02   & ALFOSC  \\
2021/06/11 & 59376.02 &    -- & 21.65     0.14   &  19.19       0.05 &  18.57  0.03  & 18.23  0.05   & ALFOSC  \\
2021/06/17 & 59382.91 &    -- &   --     &  20.70       0.16 &  19.96  0.09  & 19.54  0.08   & ALFOSC  \\
2021/06/23 & 59388.90 &    -- &   --     &  20.83       0.13 &  20.09  0.07  & 19.58  0.06   & ALFOSC  \\
2021/07/02 & 59397.17 &    -- &   --     &  21.06       0.19 &  20.25  0.08  & 19.69  0.06   & ALFOSC  \\
2021/07/10 & 59405.23 &    -- &   --     &  21.22       0.08 &  20.30  0.05  & 19.76  0.16   & ALFOSC  \\

 & & & & & & &    \\
 \hline
\end{tabular}
\end{table*}

\begin{table*}
\centering
\caption{Photometric data in the Johnson filters collected for SN 2021aai (Vega mag).}
\label{tab6}
\begin{tabular}{ c c c c c  } 
 \hline
 Date & MJD & B & V & Instrument  \\ 
 \hline
 \hline
 & & & &   \\
2021/01/19  &  59233.91 &    19.22    0.04  &  18.61    0.03   &  LRS       \\
2021/01/22  &  59236.89 &    19.32    0.07  &  18.65    0.03   &  ALFOSC    \\
2021/01/23  &  59237.08 &    19.38    0.08  &  18.61    0.07   &  LCO       \\
2021/01/24  &  59238.95 &    19.46    0.07  &  18.64    0.04   &  ALFOSC    \\
2021/01/25  &  59239.32 &    19.43    0.12  &  18.63    0.09   &  LCO       \\
2021/01/27  &  59241.24 &    19.71    0.18  &  18.62    0.12   &  LCO       \\
2021/01/31  &  59245.22 &    19.87    0.15  &  18.76    0.08   &  LCO       \\
2021/02/03  &  59248.15 &    20.16    0.08  &  18.67    0.04   &  LCO       \\
2021/02/07  &  59252.18 &    20.32    0.08  &  18.63    0.04   &  LCO       \\
2021/02/11  &  59256.23 &    20.44    0.10  &  18.67   0.03   &  LCO       \\
2021/02/20  &  59265.03 &    20.59    0.18  &  18.61    0.10   &  ALFOSC    \\
2021/02/26  &  59271.22 &     --   &  18.57    0.10   &  LCO       \\
2021/03/04  &  59277.11 &    20.81    0.11  &  18.68    0.02   &  LCO       \\
2021/03/18  &  59291.26 &    20.79    0.08  &  18.69    0.02   &  LCO       \\
2021/03/24  &  59297.23 &     --   &  18.64    0.06   &  LCO       \\
2021/05/08  &  59342.89 &    20.83    0.09  &  18.32    0.04   &  ALFOSC    \\
2021/05/09  &  59343.12 &     --   &  18.38    0.03   &  LCO       \\
2021/05/31  &  59365.88 &     --   &  18.63    0.03   &  Moravian  \\
2021/06/06  &  59371.89 &    22.37    0.09  &  19.15    0.02   &  ALFOSC    \\
2021/06/11  &  59376.01 &     --   &  20.80    0.15   &  ALFOSC    \\

 & & & &     \\
 \hline
\end{tabular}
\end{table*}

\begin{table*}
\centering
\caption{Photometric data collected through the ATLAS survey for SN 2021aai (AB mag).}
\label{tab7}
\begin{tabular}{ c c c c c  } 
 \hline
 Date & MJD & cyan  & orange  & Instrument   \\ 
 \hline
 \hline
 & & & &    \\
2021/01/24  &   59238.48  &      --    &   18.15   	    0.46  &	ATLAS	 \\
2021/01/28  &   59242.48  &      --    &   18.16   	    0.44  &	ATLAS	 \\
2021/01/30  &   59244.47  &      --    &   18.09   	    0.14  &	ATLAS	 \\
2021/02/01  &   59246.45  &      --    &   18.23   	    0.24  &	ATLAS	 \\
2021/02/05  &	59250.43  &	18.70	  0.37	 &   --   &   ATLAS	\\
2021/02/07  &	59252.45  &	18.66	  0.42	 &   --   &   ATLAS	\\
2021/02/09  &	59254.38  &	18.84	  0.30	 &   --   &   ATLAS	\\
2021/02/11  &	59256.41  &	18.74	  0.04	 &   --   &   ATLAS	\\
2021/02/13  &	59258.40  &	18.60	  0.21	 &   --   &   ATLAS	\\
2021/02/21  &   59266.49  &     18.35	  0.27   &   18.00 	    0.18  &	ATLAS	 \\
2021/02/25  &   59270.39  &      --   &    17.93   	    0.26  &	ATLAS	 \\
2021/03/03  &   59276.55  &      --    &   18.05   	    0.37  &	ATLAS	 \\
2021/03/05  &   59278.43  &      19.06	  0.16    &   17.73   	    0.20  &	ATLAS	 \\
2021/03/11  &	59284.36  &	18.67	  0.21	 &   --   &   ATLAS	\\
2021/03/15  &	59288.32  &	18.60	  0.12  &   -- 	 &   ATLAS     \\
2021/03/19  &   59292.27  &      --    &   18.11   	    0.11  &	ATLAS	 \\
2021/03/25  &   59298.34  &      --    &   17.89   	    0.43  &	ATLAS	 \\
2021/03/27  &   59300.38  &      --    &   17.67   	    0.32  &	ATLAS	 \\
2021/03/31  &   59304.29  &      --     &  17.74   	    0.28  &	ATLAS	 \\
2021/04/02  &   59306.42  &      --     &  17.85   	    0.21  &	ATLAS	 \\
2021/04/06  &	59310.28  &	18.53	  0.14  &   -- 	 &   ATLAS     \\
2021/04/10  &	59314.26  &	18.76	  0.47  &   -- 	 &   ATLAS     \\
2021/04/14  &	59318.26  &	18.47	  0.15  &   -- 	 &   ATLAS     \\
2021/04/16  &   59320.25  &      --    &   17.68   	    0.13  &	ATLAS	 \\
2021/04/18  &   59322.26  &      --    &   17.74   	    0.08  &	ATLAS	 \\
2021/04/20  &   59324.31  &      --    &   17.64   	    0.26  &	ATLAS	 \\
2021/04/24  &   59328.31  &      --    &   17.49   	    0.08  &	ATLAS	 \\
2021/04/28  &   59332.32  &      --    &   17.65   	    0.63  &	ATLAS	 \\
2021/04/30  &   59334.29  &      --    &   17.52   	    0.04  &	ATLAS	 \\
2021/05/12  &   59346.26  &      --    &   17.49   	    0.09  &	ATLAS	 \\
2021/05/14  &	59348.26  &	18.65	  0.29  &   -- 	 &   ATLAS     \\
2021/05/16  &   59350.25  &      --    &   17.61   	    0.07  &	ATLAS	 \\
2021/05/18  &   59352.26  &      --    &   17.58   	    0.42  &	ATLAS	 \\
2021/05/20  &   59354.25  &      --    &   17.51   	    0.20  &	ATLAS	 \\
2021/05/22  &   59356.26  &      --    &   17.54   	    0.07  &	ATLAS	 \\
2021/05/30  &   59364.26  &      --    &   17.53   	    0.19  &	ATLAS	 \\
2021/06/07  &   59372.26  &      --    &   18.03   	    0.19  &	ATLAS	 \\
2020/12/17  &	59200.47  &	>20.11    &   -- 	 &   ATLAS     \\
2020/12/19  &	59202.62  &	>20.08     &   -- 	 &   ATLAS     \\
2020/12/25  &   59208.44  &      --    &   >19.48  	  &	ATLAS	 \\
2020/12/29  &   59212.58  &      --    &   >18.86   &	ATLAS	 \\
2020/12/31  &   59214.53  &      --    &   >19.13   &	ATLAS	 \\
2021/01/06  &   59220.49  &      --    &   >19.76    &	ATLAS	 \\
2021/08/14  &   59440.56  &      --    &   >19.59    &	ATLAS	 \\

 & & & &    \\
 \hline
\end{tabular}
\end{table*}

\begin{table*}
\centering
\caption{Photometric data collected in the NIR for SN 2021aai (Vega mag).}
\label{tab8}
\begin{tabular}{ c c c c c c  } 
 \hline
 Date & MJD & J  & H & K &  Instrument   \\ 
 \hline
 \hline
 & & & & &   \\
2021/01/25  &  59239.93  &  17.24    0.19 & 16.90  0.04 & 16.66  0.09 & NOTCAM \\
2021/02/18  &  59263.96  &  16.84    0.12 & 16.42  0.12 &  --   & NOTCAM \\
2021/05/22  &  59356.88  &  16.11    0.12 & 15.81  0.05 & 15.60  0.13 & NOTCAM \\
 & & & & &  \\
 \hline
\end{tabular}
\end{table*}

\bsp	
\label{lastpage}
\end{document}